\documentclass[fleqn,twoside]{article}
\usepackage{espcrc2}

\usepackage{graphicx}

\usepackage{epsfig}

\usepackage[figuresright]{rotating}

\newcommand{\AmS}{{\protect\the\textfont2
  A\kern-.1667em\lower.5ex\hbox{M}\kern-.125emS}}

\title{
The Spin Structure of the Proton and Polarized Collider Physics}

\author{S.D. Bass\address[ECT]{ECT*, Strada delle Tabarelle 286, 
        I-38050
        Villazzano, Trento, Italy} and
        A. De Roeck \address[CERN]{CERN, CH-1211 Geneve 23, Switzerland} }

\begin{document}

\begin{abstract}
We summarise the present status of the proton spin problem and the 
physics possibilities for future polarized $ep$ and $pp$ colliders.
This summary is based on the presentations and discussion sessions
at the workshop 
``The Spin Structure of the Proton and Polarized Collider Physics'' 
(Trento, July 23-28, 2001).
\vspace{1pc}
\end{abstract}

\maketitle

\section{INTRODUCTION}

Understanding the internal spin structure of the proton is one of the most 
challenging problems facing particle physics today.
How is the spin of the proton built up from the intrinsic spin and orbital 
angular momentum of its quark and gluonic constituents ?
What happens to spin and orbital angular momentum in the transition between 
current and constituent quarks in low-energy QCD ?

The small value of the flavour-singlet axial charge $g_A^{(0)}$ extracted 
from the first moment of $g_1$ 
(the nucleon's first spin dependent structure function) 
\begin{equation}
\left. g^{(0)}_A \right|_{\rm pDIS} = 0.2 - 0.35
\end{equation}
in 
polarized fixed target deep inelastic scattering experiments has inspired 
vast activity on both the theoretical and 
experimental sides to understand the internal spin structure of the proton
-- for recent reviews see \cite{windmolders,mauro,bass99,reya,sofferrhic}.

In relativistic quark models and in the naive (pre-QCD) parton model the 
flavour-singlet axial charge $g_A^{(0)}$ is interpreted as the fraction 
of the proton's spin which is carried by its quark and anti-quark (partonic) 
constituents.
Relativistic quark models generally predict that $g_A^{(0)}$ 
should be approximately equal to the Ellis-Jaffe \cite{ej} 
(or OZI) value, about 0.6, implying the questions: 
where is the ``missing'' spin and what is the QCD physics of $g_A^{(0)}$ ?

The (initial EMC \cite{emc}) measurement of $g^{(0)}_A|_{\rm pDIS}$, 
one of the main ``surprises'' in recent high-energy physics,
%
has inspired about 1000 theoretical papers and a new programme of 
dedicated experiments: 
high-energy polarized deep inelastic scattering at CERN \cite{kunne}, 
DESY \cite{fantoni}, SLAC, and polarized proton-proton collisions at 
RHIC \cite{saito,kurita}, 
as well as lower-energy experiments at Jefferson Laboratory 
\cite{meziani}, ELSA and MAMI \cite{helbing}  to understand 
the spin structure of the proton. 
Further proposals aimed at completing a definitive study exist
for a polarized $ep$ collider \cite{deroeck}
(polarized HERA collider at DESY \cite{abhay}
 or EIC collider at BNL \cite{deshpande,krasny})
and elastic $\nu p$ scattering \cite{tayloe}.
The proton spin problem has also inspired new theoretical 
thinking and experimental activity on the famous $U_A(1)$ problem of QCD.

In this summary of the workshop
``The Spin Structure of the Proton and Polarized Collider Physics''
(Trento, July 23-28, 2001)
we review the key physics issues and experimental status.
We start in this Section with a brief theoretical overview, following in
Section 2 with an overview of the present experimental situation and 
future possibilities.
The key measurements to disentangle to the proton's internal 
spin structure are discussed in more detail in the Sections after that.

Much theoretical work has revealed that, in QCD, $g_A^{(0)}$ receives
contributions from quark and gluon partons \cite{ar,et} and a possible 
non-perturbative contribution at $x=0$ from gluon topology \cite{bass}:
\begin{equation}
g_A^{(0)} 
= 
\Biggl(
\sum_q \Delta q - 3 {\alpha_s \over 2 \pi} \Delta g \Biggr)_{\rm partons}
+ \ {\cal C}
\end{equation}
Here ${1 \over 2} \Delta q$ and $\Delta g$ are the amount of spin carried 
by quark and gluon partons in the polarized proton and ${\cal C}$ is the 
topological contribution.

Hence the key driving question is to pin down experimentally 
the size of each of the different contributions and 
to resolve the spin-flavour structure of the proton.
This is the main goal of the present and next generation of experiments.
So far the main experimental activity has focused on fully 
inclusive measurements of the proton's $g_1$ spin structure 
function with longitudinally polarized targets.
The key issues for the new experiments are to measure the 
separate flavour- and spin-dependent parton distributions 
for the proton's valence, sea quark and gluonic constituents,
and to investigate the spin structure of transversely polarized
protons.
We now briefly discuss the different contributions;
further details on 
present status and measurability are given in the Sections below.

Physicswise, $\Delta q_{\rm partons}$ is associated with the hard 
photon scattering on quarks and antiquarks with 
low transverse momentum squared, $k_t^2$ of the order of typical 
gluon virtualities in the proton, and 
$\Delta g_{\rm partons}$ is associated with the 
hard photon scattering on quarks and antiquarks carrying $k_t^2 \sim Q^2$
\cite{ccm,bint}.
The partonic decomposition (2) holds also in 
the AB \cite{ab} and JET \cite{jet} factorization schemes;
in the ${\overline {\rm MS}}$ factorization scheme \cite{msbar} 
the full partonic contribution is written as
$\Delta q_{\overline {\rm MS}} =
(
\Delta q - {\alpha_s \over 2 \pi} \Delta g )_{\rm partons}$.
The product $\alpha_s \Delta g$ in (2) scales 
when $Q^2 \rightarrow \infty$ under QCD evolution 
\cite{ar}. 
This means that gluon polarization plays a potentially key 
role in any understanding of the proton spin puzzle.
One possible explanation of the proton spin problem is that 
$(\Delta u + \Delta d + \Delta s)_{\rm partons}$
takes approximately 
the Ellis-Jaffe value, about 0.6, and
that
(in the absence of the topology term - see below)
the OZI suppression of 
$g_A^{(0)}|_{\rm pDIS}$
is induced by large gluon polarization 
$\Delta g \sim 2$ at $Q^2 \sim 1$GeV$^2$.
Measuring the size of $\Delta g$ is the key issue 
for high-energy investigations of the proton's internal spin structure.

QCD motivated fits have been applied to the existing $g_1$ data set to 
try to deduce information about the size of $\Delta g$.
The main source of error in these QCD fits comes from lack of knowledge 
about $g_1$ in the small $x$ region and (theoretical) the functional 
form chosen for the quark and gluon distributions in the fits 
\cite{abfr,def,smcqcd}.
Precise data from a future polarized $ep$ collider at low $x$ would 
be extremely valuable here.
Various approaches are planned aiming at a direct measurement of 
$\Delta g$: 
through semi-inclusive measurements of charm and high $p_t$ 
hadron pair production in polarized leptoproduction 
(COMPASS and HERMES) and photoproduction (SLAC) experiments, 
and studies of prompt photon 
production in polarized proton-proton collisions at RHIC 
-- see Section 4 below.
In the longer term one would like to make a decisive measurement 
via  study of two-quark jet events in $\gamma^* g$ fusion at a 
future polarized $ep$ collider (polarized HERA or EIC at BNL) \cite{radel}.

The flavour dependence of the $\Delta q_{\rm partons}$ may be probed
through semi-inclusive measurements in the current fragmentation 
region of 
(e.g.) 
fast pions in the final state of polarized leptoproduction experiments.
Modulo assumptions on the fragmentation, these data may be used 
to reconstruct the polarized valence and sea distributions.
More direct measurements will come from $W$ boson production in polarized
$pp$ collisions at RHIC and from charged current exchange processes in
high $Q^2$ deep inelastic scattering.

The topological contribution ${\cal C}$ in (2) is associated with 
Bjorken $x$ equal to zero and is related to long range gluon dynamics 
\cite{bass}.
Polarized deep inelastic scattering experiments cannot access $x=0$ and
measure
$g_A^{(0)}|_{\rm pDIS} = g_A^{(0)} - {\cal C}$. 
The full $g_A^{(0)}$ may be measured through elastic $Z^0$ exchange in
$\nu p$ elastic scattering.
A finite value of ${\cal C}$ is natural with spontaneous 
$U_A(1)$ symmetry 
breaking by instantons 
where any instanton induced suppression of $g_A^{(0)}|_{\rm pDIS}$ 
(the axial charge carried by partons with finite momentum fraction
 $x>0$)
is compensated by a shift of axial charge to the zero-mode so that
the total axial-charge $g_A^{(0)}$ including ${\cal C}$ is conserved.
An overview of instanton physics is given in \cite{marga}.
A quality measurement of $\nu p$ elastic scattering would provide very 
valuable information about the nature of $U_A(1)$ symmetry breaking in 
QCD.
If some fraction of the spin of the constituent quark is carried by 
gluon topology in QCD (at $x=0$),  then the constituent quark model 
predictions for $g_A^{(0)}$ are not necessarily in contradiction with 
the small value of $g_A^{(0)}|_{\rm pDIS}$ extracted from deep inelastic 
scattering experiments.
A decisive $\nu p$ elastic experiment may be possible using the miniBooNE
experiment at FNAL \cite{tayloe}.

In the isotriplet channel things are much better understood.
The fundamental Bjorken sum-rule \cite{bj}
\begin{equation}
\int_0^1 dx (g_1^p - g_1^n)(x,Q^2) = {1 \over 6} g_A^{(3)} C_{NS}(Q^2)
\end{equation}
is experimentally verified to 10\% accuracy by present fixed target
experiments.
Here $g_A^{(3)} = (\Delta u - \Delta d)_{\rm partons}$ is the proton's 
isotriplet axial-charge which is measured also in neutron $\beta$ decay 
and $C_{NS}(Q^2) = (1 + O({{\alpha_s}\over  \pi}))$ is the perturbative 
QCD correction.
The Bjorken sum-rule is a rigorous prediction of current algebra and QCD.
The Goldberger-Treiman relation from chiral symmetry,
$2m g_A^{(3)} = f_{\pi} g_{\pi NN}$,
relates 
$(\Delta u - \Delta d)$ to the pion-nucleon coupling constant 
$g_{\pi NN}$ 
extracted from dispersion relation analyses of low-energy 
$\pi N$ scattering,
meaning that the spin structure of the proton measured in
high-energy polarized deep inelastic scattering experiments
is intimately related to spontaneous chiral symmetry breaking in QCD.

The interplay between the proton spin problem and the $U_A(1)$ 
problem is further manifest in the flavour-singlet 
Goldberger-Treiman relation \cite{venez} 
which connects $g_A^{(0)}$ with the $\eta'$--nucleon coupling constant 
$g_{\eta' NN}$.
Working in the chiral limit it reads
\begin{equation}
m g_A^{(0)} = \sqrt{3 \over 2} F_0 \biggl( g_{\eta' NN} - g_{QNN} \biggr) 
\end{equation}
where
$g_{\eta' NN}$ is the $\eta'$--nucleon coupling constant and $g_{QNN}$ 
is an OZI violating coupling which measures the one particle irreducible 
coupling of the topological charge density 
$Q = {\alpha_s \over 4 \pi} G {\tilde G}$ 
to the nucleon
($m$ is the nucleon mass and $F_0$ $\sim 0.1$GeV 
 renormalises the flavour-singlet decay constant).
There is presently a vigorous experimental programme aimed at studying 
the low-energy $\eta'$-nucleon interaction, 
including possible OZI violations, at COSY and Jefferson Laboratory.

The shape of $g_1$ is particularly interesting.
Valuable information about the internal valence structure of the nucleon 
will come from studying the large $x$ region (close to one) where SU(6) 
and perturbative QCD predict that the ratio of polarized to unpolarized 
structure functions for both proton and neutron should go to one when 
$x \rightarrow 1$ \cite{athomas}.
Precision measurements of large $x$ spin asymmetries will be possible 
following the upgrade of Jefferson Laboratory to 12 GeV \cite{meziani}.
Small $x$ measurements \cite{badelek} from polarized $ep$ colliders
would provide valuable information about perturbative QCD dynamics at 
low $x$, where the shape of $g_1$ 
is particularly sensitive to the effects of
$(\alpha_s \ln^2 {1 \over x})^k$ 
resummation and DGLAP evolution \cite{kwiecinski}.

Spin transfer reactions also have the potential to provide valuable 
insight into the role of spin in QCD hadronization dynamics
\cite{soffer}.
To obtain a leading twist spin-transfer asymmetry 
requires measurement of the polarization of one of the 
outgoing particles in addition to having a polarized beam or target.
The self-analysing properties of the $\Lambda$ hyperon through 
its dominant weak decay $\Lambda \rightarrow p \pi^-$ 
make it particularly appealing for experimental study; 
recent studies of $\Lambda$ production at LEP have demonstrated 
the feasibility of successfully reconstructing the $\Lambda$ spin
\cite{leplambda}.
The fragmentation functions $D_f^{H}(z,Q^2)$ which describe
the non-perturbative hadronization process are the timelike 
counterpart of the usual parton distribution functions measured in 
spacelike deep inelastic scattering and represent the probability 
to find the hadron $H$ with fraction $z$ of the momentum of the 
parent parton $f$ at a given value of $Q^2$.
They are expected to satisfy perturbative QCD factorization and 
to evolve under QCD evolution equations.
To date the best known fragmentation functions are those for the
most copiously produced light mesons $\pi$ and $K$.
Measurement of the $z$ dependence of semi-inclusive polarized 
$\Lambda$ production 
in polarized deep inelastic scattering at HERMES and COMPASS 
will help to resolve different models of the fragmentation functions.
The (V-A) structure of charged current exchange in deep inelastic 
scattering provides a source of polarized quarks with specific 
flavour \cite{soffer}.
Semi-inclusive $\Lambda$ production events at 
(unpolarized) HERA
could be analyzed to deduce valuable information on the polarized
fragmentation functions.
At the high energy and luminosity of RHIC
the
rapidity distribution of longitudinally produced
$\Lambda$'s in 
single-spin 
${\vec p}p \rightarrow {\vec \Lambda} X$ collisions 
is
particularly sensitive to the spin dependent structure of the
$\Lambda$ fragmentation.

Further experimental study of baryon fragmentation is also being
actively investigated through semi-inclusive measurements of 
final state hadron multiplicities in deep inelastic scattering 
from nuclear targets using the RICH detector at HERMES 
\cite{muccifora}.
These data may provide some insight into the hadron formation times.
The present HERMES data from nitrogen and krypton targets show 
no significant difference between the multiplicity ratios 
for $\pi^+$ and $\pi^-$ production, 
which are significantly greater 
than the multiplicity ratio for positively charged hadron production
for both targets. These data have been interpreted \cite{muccifora} 
to mean that the proton has a longer formation time than a pion.

In polarized photoproduction the Gerasimov-Drell-Hearn sum-rule 
\cite{gerasimov} provides a further important constraint on the 
spin structure of the proton.
The GDH sum-rule relates the difference of the two cross-sections 
for the absorption of a real photon with spin anti-parallel 
$\sigma_A$ and parallel $\sigma_P$ to the target spin to the 
square of the anomalous magnetic moment of the target, viz.
\begin{equation}
\int_{\nu_{th}}^{\infty} {d \nu \over \nu} (\sigma_A - \sigma_P)(\nu)
= - {2 \pi^2 \alpha \kappa^2 \over m^2} 
\end{equation}
Here $m$ is the mass of the target; $\kappa$ is the proton's anomalous 
magnetic moment.
The GDH sum-rule is derived from general principles 
(see section 6 below)
and any measured violation would yield new 
challenges for theorists to explain the QCD dynamics of such an effect.
Dedicated real-photon experiments at ELSA, MAMI \cite{helbing}
and SLAC are presently
planned or underway to test the GDH sum-rule.
The low-energy part is dominated by spin structure in the resonance 
region; the high-energy part is expected to be dominated by spin-dependent 
Regge dynamics.

Exclusive measurements in deeply virtual leptoproduction have recently
attracted much attention.
Deeply virtual Compton scattering (DVCS) provides a 
possible experimental 
tool to access the quark total angular momentum, $J_q$, 
in the proton through generalized parton distributions (GPDs) \cite{ji}.
The HERMES experiment \cite{amarian}
measure in the kinematics where they expect to be dominated by 
the DVCS-Bethe-Heitler
interference term and observe the $\sin \phi$ 
azimuthal angle and 
helicity dependence expected for this contribution.
At lower energies virtual Compton scattering experiments at 
Jefferson Laboratory and MAMI probe the electromagnetic deformation of 
the proton and measure the electric and magnetic polarizabilities of the 
target \cite{disalvo}.

The study of single spin asymmetries from transversely polarized 
targets is currently a subject of much interest 
\cite{matthias,boer,anselmino,drago}.
These asymmetries are expected to yield information about the 
density of transversely polarized quarks inside a transversely 
polarized proton.
The difference between the transversity and helicity distributions
reflects the relativistic character of quark motion in the nucleon
\cite{jaffe01}.

The HERMES experiment at DESY is expected to start measurements with
transverse target polarization in the near future.
The observable is Collins' single spin asymmetry \cite{collins}
for charged pion production in deep inelastic electron proton scattering.
The azimuthal distribution of the final state pions with respect 
to the virtual photon axis carries information about the transverse 
quark spin orientation. 
Further possibilities to access transversity 
include proposed measurements of 
transverse single spin asymmetries in $pp$ collisions at RHIC and 
via interference fragmentation functions extracted from 
light quark di-jet production in $e^+ e^-$ collisions at LEP or
B-factories.

Finally, spin asymmetries offer new variables to search for new physics
beyond the Standard Model, or, when found, to study the chiral structure
of these new interactions. Already at polarized RHIC there are several 
examples of new physics searches and studies demonstrating the power of 
polarized beams. Therefore it would be useful to keep such a possibility 
also in mind for what one hopes will become the ``parade horse'' to study 
new physics, namely the Large Hadron Collider at CERN. 
If the new physics warrants it, polarization  could be considered as an 
upgrade of this machine.

\section{EXPERIMENTAL FACILITIES}
\vspace{1mm}
\noindent

Most of the data on polarized scattering to date is provided
by deep inelastic scattering (DIS) experiments. 
Mirrored to the unpolarized scattering experiments, polarized 
lepton beams and polarized fixed targets have been employed.
A pioneering experiment
was the Yale-SLAC collaboration~\cite{yale}, 
which measured DIS down to
$x= 0.1$. The measurements of this experiment were consistent 
with the naive parton model view that 
$\sim 60\%$ of the nucleon spin
is carried by its quark parton constituents.
The EMC experiment~\cite{emc} at CERN used a muon beam and 
extended the inclusive measurements down to
$x = 0.01$. 
The EMC data led to the spin puzzle, hinting 
that the contribution of the quarks to the spin of the proton is small.
These findings 
triggered a 
whole program on DIS fixed target
experiments, 
with electron beams at SLAC~\cite{slac}, 
a muon beam (SMC)~\cite{kunne} at CERN and 
the electron ring of the HERA collider on an internal target 
(HERMES)~\cite{fantoni} at DESY. 
The latter uses the self-polarization of the electron beam
via the Sokolov-Ternov effect~\cite{sokolov}
to achieve polarized electron beams with 
about 60\% polarization.
The lowest values in $x$ for a $Q^2$ around 1 GeV$^2$
which can be achieved in these experiments 
amounts to approximately $10^{-3}$ in SMC, due to the high energy of the 
muon beam.

All of these experiments, apart from HERMES, have by now been concluded.
They confirmed the spin puzzle with much increased precision, and launched 
a series of new measurements, such as semi-inclusive ones.
Since a few years, the possible culprit to the spin puzzle is thought 
to be the polarized gluon distribution, which could be large. 
Hence a new fixed target experiment COMPASS \cite{kunne}
has been assembled in the last years, at the muon beam at CERN, 
with its main mission to make a direct measurement of $\Delta g$ 
in the region $x \sim 0.1$. 
This experiment will start data taking in 2002.

At SLAC the experiments E159 and E161 are scheduled, and 
will use polarized real-photon nucleon interactions for spin studies. 
The former will study in particular the 
GDH sum rule, while the latter plans to measure $\Delta g$ from open
charm production.

Meanwhile it was realised that polarized $pp$ collisions could also 
give information on the polarized gluon, and the $pp$ and heavy-ion 
collider RHIC at BNL was 
`upgraded'
within its construction phase 
to include the option of polarized $pp$ scattering \cite{saito}.
Polarization of the proton beam is technically more involved 
than for the electron beam, since protons do not 
polarize naturally in a storage ring, at least not within any 
useful time span.
Hence beams from a source of polarized protons have to be
accelerated through the whole chain, keeping the 
polarization during the process. Special magnets called 
Siberian Snakes allow the protons to pass depolarizing resonances 
during the acceleration and correct for depolarizing distortions 
during the storage of the beam.
Thus polarized $pp$ collisions with a centre of mass system (CMS)
energy in the range of
200 and 500 GeV could be provided at RHIC, with integrated luminosities
of 320 and 800 pb$^{-1}$ respectively.

On the DIS front, 
the commissioning of the HERA electron-proton collider 
(27.5~GeV electrons on 820~GeV protons)
eight
years ago opened up a completely new kinematical domain 
in deep inelastic scattering (DIS), and the two HERA experiments have provided 
a multitude of new insights into the unpolarized structure of the 
proton and the photon since then. 

The success of HERA has prompted thinking about future $ep$ 
collider opportunities for polarized scattering or $eA$ collisions. 
Possible high energy $ep$ collider projects 
presently under discussion are listed in Table~\ref{tab1}.

\begin{table}
\caption{Possible future $ep$-collider facilities for polarized scattering}
\label{tab1}
\begin{tabular}{|l|l|l|}
\hline
Machine  & Lumi/year  & $\sqrt{s}$ \\
\hline
HERA   & 150 pb$^{-1}$& 320 GeV \\
Electron-Ion & $ 4$ fb$^{-1}$ & 30-100 GeV\\
Collider (EIC) & & \\
THERA  & 40 (250) pb$^{-1}$ &  1-1.6 TeV \\
(TESLA$\otimes$HERA) & & \\
\hline
\end{tabular}
\end{table}

All these projects are in different stages of development. 
HERA is in principle closest to reaching this goal. 
The accelerator exists and has just been upgraded
to reach the luminosity quoted. 
It has well understood detectors, and the electron beam is 
already polarized.
The polarization of the proton beam can probably be achieved 
in the same way as for RHIC.
The technical aspects of this project are elaborated in~\cite{barber} and
physics studies reported in \cite{gehr,works}. 
Based on these studies, it seems realistic to assume that 
HERA could be operated with polarized electron and proton beams, each  
polarized to about 70\%, reaching a luminosity of 
500~pb$^{-1}$  when integrated over several years. 
Note that the polarized protons from HERA could also 
be used to collide with a polarized internal target \cite{heran}, 
producing polarized $pp$ 
collisions, as at RHIC, but at a much reduced CMS energy.

The Electron Ion Collider (EIC)~\cite{erhic}
-- if built at BNL -- will need a polarized 
electron accelerator, either a ring or  LINAC, added to the RHIC polarized
proton rings, and will  probably also need a dedicated
experiment. 
While an interesting program has been developed for the lower energy 
end of the EIC, in this paper we will usually refer to it as a machine with 
a CMS energy of 100 GeV.
The advantage of the EIC
is its large reachable luminosity, imperative for 
polarized studies. 
At HERA the luminosity is  (just) enough for most 
topics but its advantage lies in its larger kinematical reach.

It would be very advantageous to have polarized low-$x$ neutron data at 
future $ep$ colliders, which would enable measurement of both singlet 
and non-singlet polarized structure functions at low $x$.
A study was made of the potential for using polarized $^3$He in 
HERA~\cite{abhay}, 
which would enable $g_1^n$ to be extracted. 
If the machine can provide polarized $^3$He
with a luminosity comparable to the one for the protons, 
such a program
can be carried out.
A  recent new idea is to store polarized deuterons~\cite{derb}.
Due to the small 
gyromagnetic anomaly
value the storage and acceleration problems are less severe
for deuterons and
it could be possible to keep the polarization at HERA
even without the use of 
Siberian Snake magnets for deuterons. However the spin
cannot
be flipped by spin rotators from transverse to longitudinal 
polarization in the interaction regions, and other means, such as the 
recently suggested use of external radial frequency fields must be considered
for arranging that the spins align longitudinally at the interaction 
points.
For polarized  deuteron beams
the changes to the present HERA machine could be more modest and
cheaper than protons, and,
 if one can instrument the region around the beam-pipe to tag the 
spectator in the deuteron nucleus, it can simultaneously give
 samples of scattering on $p$ and $n$.
An intense polarized deuteron source is however needed. 
Furthermore, so far the  acceleration and storage
of polarized deuteron beams 
has been studied to a much lesser extent than for polarized protons.

While the HERA and EIC projects could still be realised in the next 
10 years, 
ideas for even longer term projects are already being discussed. 
In particular,
several projects have been considered in connection with the 
proposal for the linear accelerator TESLA, at DESY, 
which can produce high energy polarized electron (and positron) beams.
THERA proposes to collide the 
electron beam on the protons of HERA~\cite{thera,deroeckt}.
Thus THERA will need TESLA to be built at DESY (or tangential to the 
TEVATRON), polarized protons in the proton ring, and a new detector.
THERA reaches even further in the  kinematic plane, but its
relatively low luminosity of 40 pb$^{-1}$/year 
 may be a handicap for many studies.

It is also proposed to use the polarized electron beam of TESLA for 
fixed target experiments \cite{nowak}. 
TESLA-N proposes to use beams in the energy range of 30-250 GeV on a 
(polarized) fixed target. 
ELFE@DESY proposes a 30 GeV beam, stretched in HERA to increase the 
duty cycle of the facility, on a fixed target.

Continuous 6 GeV electron beams (100\% duty cycle) are presently 
available at Jefferson Laboratory;   
the proposed upgrade \cite{meziani}
to 12 GeV 
(also 100\% duty cycle) 
would enable precision key studies of the valence structure 
of the proton in the large $x$ region (greater than 0.6),
semi-inclusive measurements of the spin-flavour structure at large 
$x$ and studies of the spin structure of transversely polarized 
nucleon targets as well as investigations of exclusive reactions at 
intermediate $Q^2$.

\begin{figure}[t!]
\begin{center}
\epsfig{file=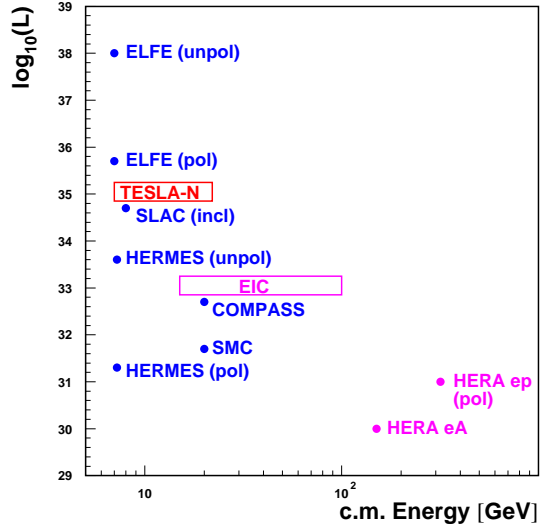,bbllx=30pt,bblly=30pt,bburx=540pt,bbury=480pt,width=8cm}
\caption{Luminosity versus CMS energy for existing and possible future 
facilities \protect\cite{nowak}.}
\label{fig:cmslumi}
\end{center}
\end{figure}

Fig.~\ref{fig:cmslumi} shows the luminosity versus CMS energy of most of 
these future 
facilities. Note that dilution factors as they appear in 
most fixed target
experiments using targets such as NH$_3$ are not taken into account here
(and would lower the luminosity of the fixed target experiments).

A neutrino factory from e.g. a 50 GeV muon beam with 
10$^{20}$ muon decays/year, would allow the use of polarized targets
for polarized $\nu p$ scattering experiments. These experiments 
would allow a plethora of new proton spin studies, in particular 
in disentangling the flavour dependence of the polarized parton
distributions. Clearly there are still quite a few technical
hurdles to be overcome before one can think of planning such a facility.

When looking into the far future, we should also include the LHC.
The LHC will be the machine with the highest energy for a long time to come
and will undergo upgrade programs (luminosity and energy) in the decade
after the start-up. Technically one would need to make room for Siberian 
Snakes and spin rotator magnets in the machine, which are at presently not 
foreseen. To our knowledge no machine study has been made yet on
how to polarize 
the LHC, but the success of RHIC in the years to come could become
infectious. A study was made for the SSC~\cite{krisch} and 
the project was considered feasible. 
As a rule of thumb one snake would be necessary for each 2 km of accelerator.
In particular if polarization would be a useful tool to study better the 
putative new physics which will open in the range of the LHC, this option 
will be very much wanted.

\begin{figure}[t!]
\begin{center}
\epsfig{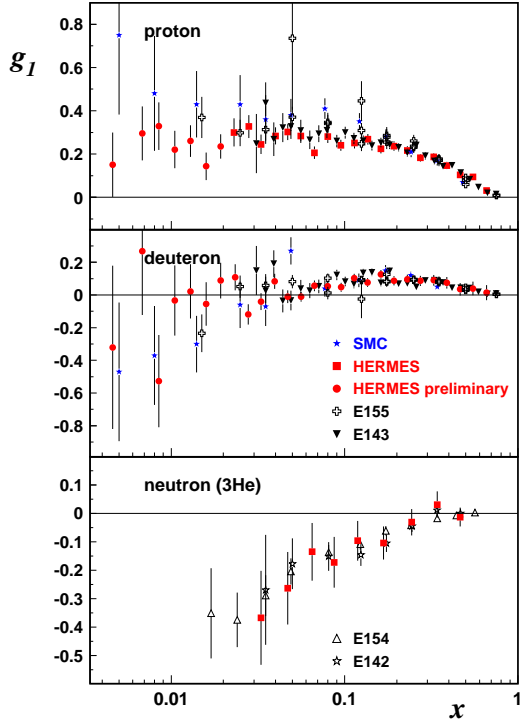}
\caption{ Compilation of $g_1$ data ($Q^2 > 
1$GeV$^2$) \protect \cite{fantoni}.}
\label{fig:g1world}
\end{center}
\end{figure}

\begin{figure}[t!]
\begin{center}
\epsfig{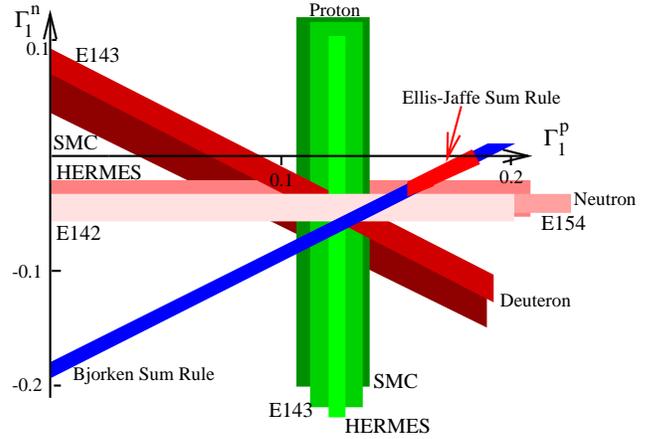}
\caption{ Compilation of
$\Gamma^p_1, \Gamma^n_1$ world data \protect \cite{fantoni}.}
\label{fig:sumrule}
\end{center}
\end{figure}


\begin{figure}[t!]
\begin{center}
\epsfig{file=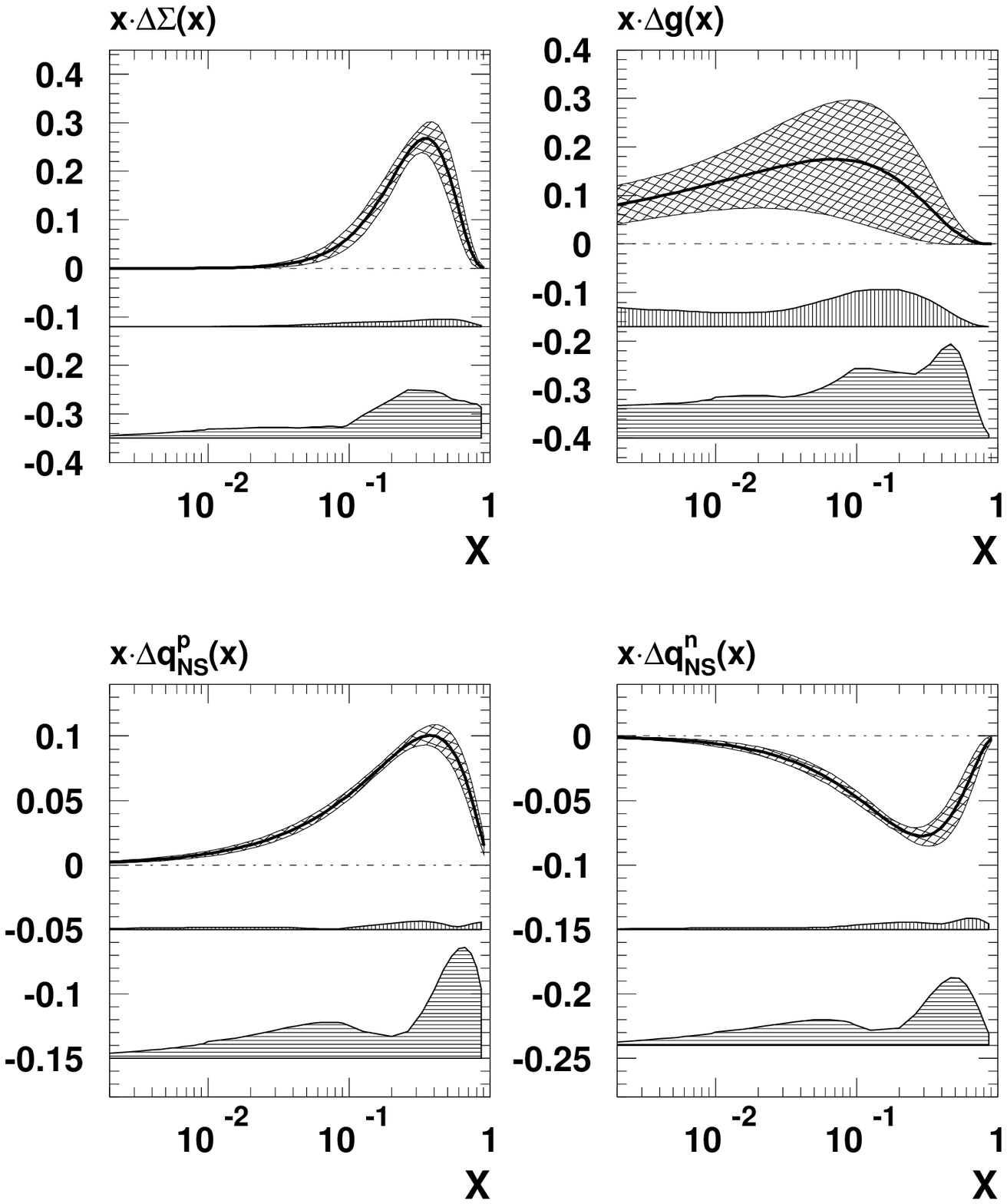,bbllx=0pt,bblly=20pt,bburx=520pt,bbury=560pt,width=8cm}
\caption{ Polarized parton distribution functions determined at 
$Q^2$ = 1 GeV$^2$ 
with statistical (upper band), experimental systematic (medium band)
and theoretical (lower band) uncertainties
 \protect \cite{kunne}.}
\label{fig:partons}
\end{center}
\end{figure}

\section{INCLUSIVE POLARIZED MEASUREMENTS}
\vspace{1mm}
\noindent

Deep inelastic structure functions are extracted from inclusive 
electron-proton or electron-nucleus scattering. 
The corresponding observable of the $F_1$ structure function in 
unpolarized scattering is the structure function $g_1$, 
which in leading-order is directly related to the polarized 
quark distributions.
Until now the quality of the polarized data cannot compete with 
the quality of the unpolarized data. 
The kinematic range is also smaller.

An overview of part of the world data on the nucleon spin structure 
is shown in Fig.~\ref{fig:g1world}. 
There is a general consistency between all data sets. 
HERMES is still collecting data.
Also, their full deuterium set is not yet analysed; 
soon $g_1^d$ will be of unprecedented statistical accuracy.

\begin{figure}[t!]
\begin{center}
\epsfig{file=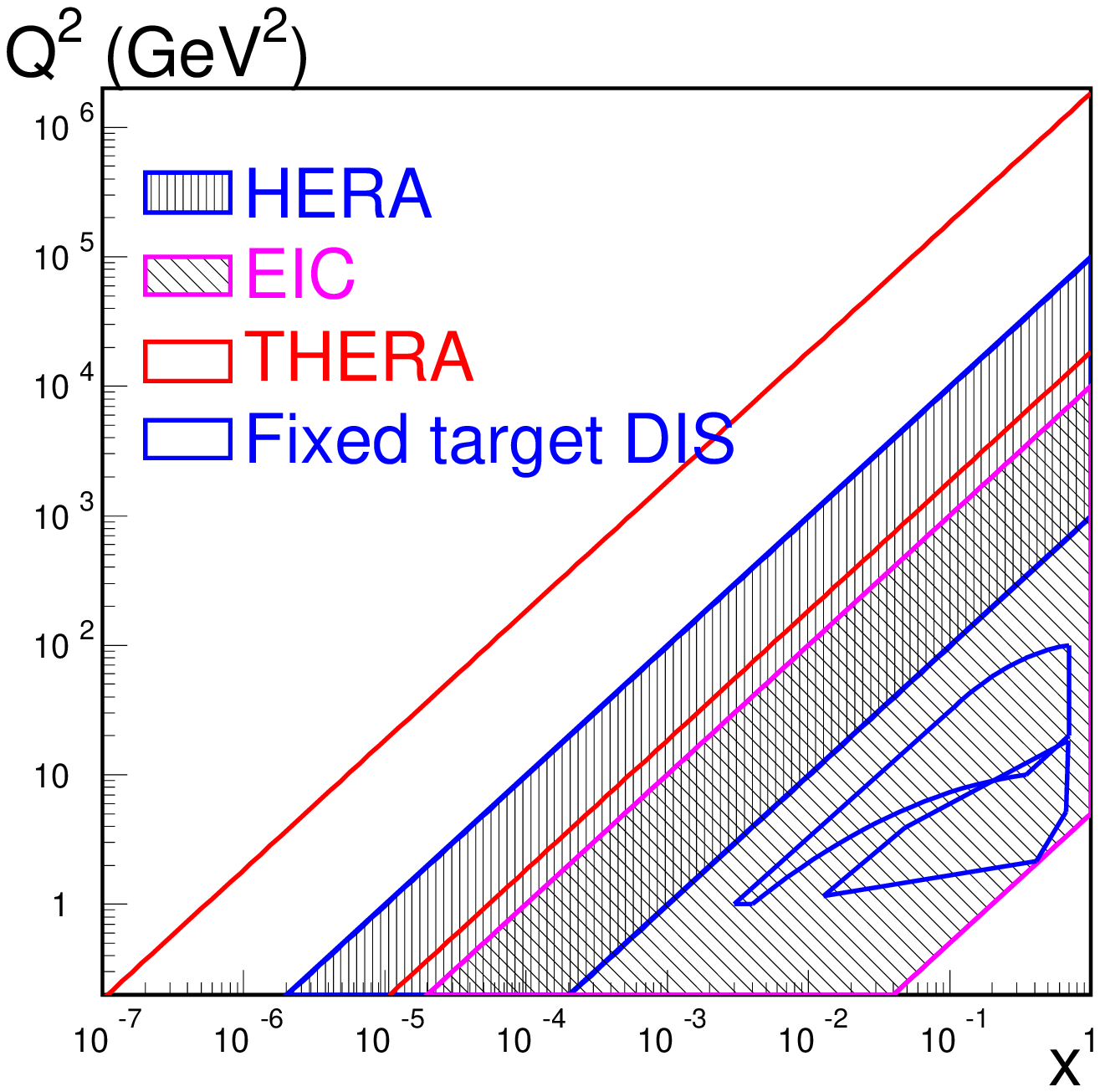,bbllx=90pt,bblly=230pt,bburx=500pt,bbury=600pt,width=8cm}
\caption{Measurable $x-Q^2$ region for a polarized HERA with the 
presently explored regions by fixed target 
experiments. The energies used are for 
THERA: $E_p = 920, E_e=500$ GeV;
  HERA: $E_p = 920, E_e=27$ GeV;
  EIC: $E_p=250, E_e=10$ GeV and 
        $ E_p=50, E_e=5$ GeV, and roughly $0.01 < y < 1.0$
 \protect \cite{abhay}.}
\label{fig:kin}
\end{center}
\end{figure}

\begin{figure}[t!]
\begin{center}
~ \epsfig{file=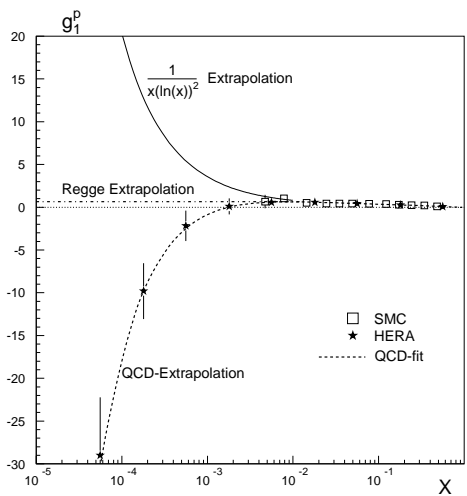,bbllx=0pt,bblly=0pt,bburx=560pt,bbury=560pt,width=7cm}
\caption{ The statistical uncertainty on the structure function $g_1$ of
the proton 
measurable at HERA, evolved to a value of 
 $Q^2 = 10$ GeV$^2$ for an integrated luminosity of
500 pb$^{-1}$. The SMC measurements are shown for comparison
 \protect \cite{deroeck}.}
\label{fig:g1p}
\end{center}
\end{figure}

\begin{figure}[t!]
\begin{center}
\epsfig{file=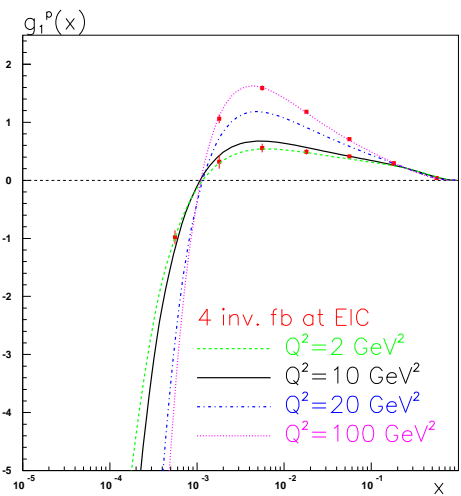,bbllx=0pt,bblly=40pt,bburx=550pt,bbury=550pt,width=7cm}
\caption{
The statistical uncertainty on the structure function $g_1$ of
the proton 
measurable at EIC,  
  for an integrated luminosity of
4 fb$^{-1}$  \protect \cite{abhay}.}
\label{fig:g1peic}
\end{center}
\end{figure}

\begin{figure}[t!]
\begin{center}
\epsfig{file=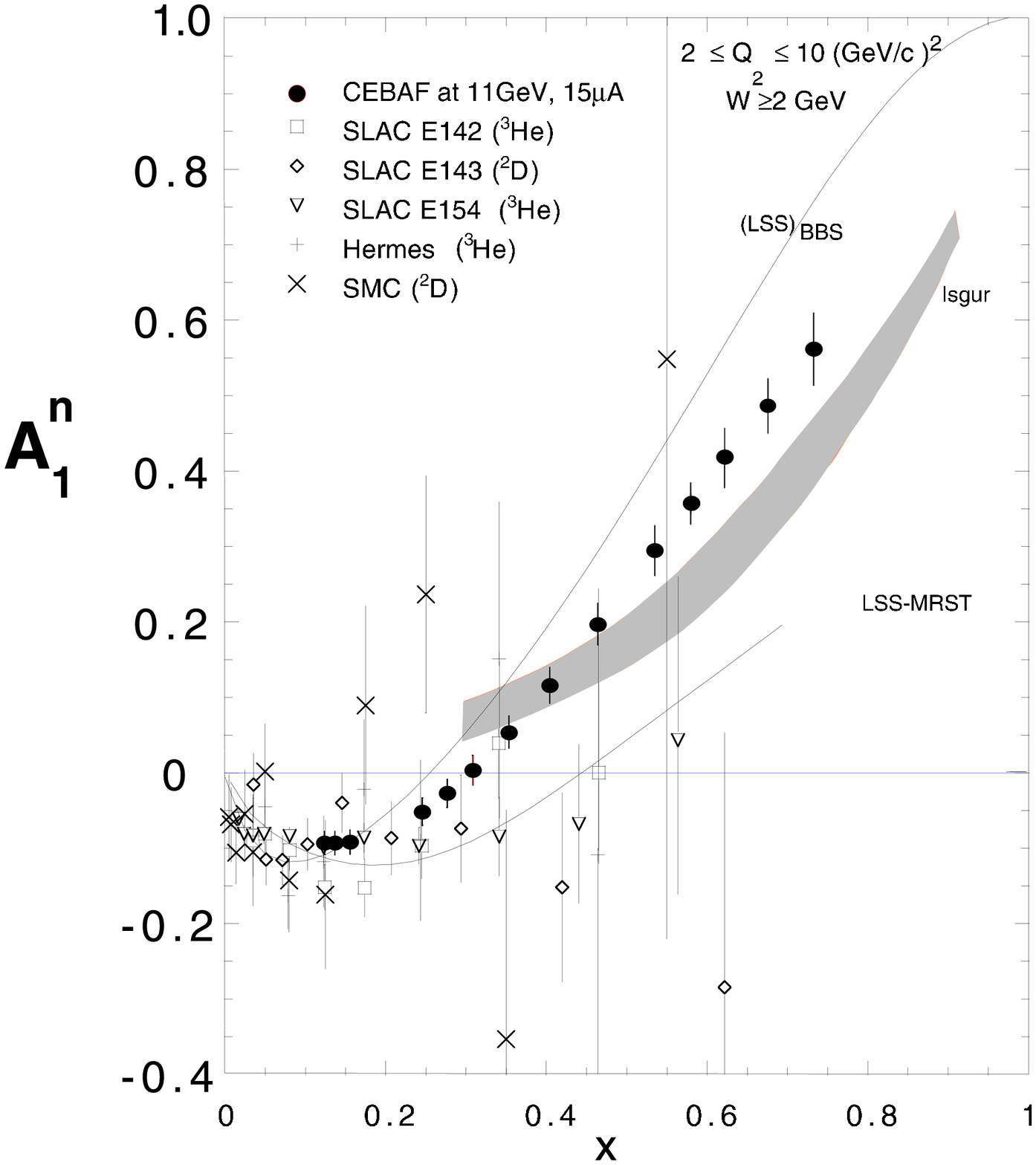,bbllx=50pt,bblly=50pt,bburx=550pt,bbury=830pt,width=6.5cm}
\caption{Projected data achievable on $A_{1n}$ using an 11 GeV polarized 
          electron beam at 
          Jefferson Laboratory \protect \cite{meziani}.}
\label{fig:jlab12}
\end{center}
\end{figure}


The largest range is provided by the SMC experiment~\cite{kunne}, namely
$0.00006 < x < 0.8$ and $0.02 < Q^2< 100$ GeV$^2$. 
This experiment
used proton and deuteron targets, with 100-200 GeV muon beams. The
final results are given in ~\cite{smcfinal}.
The low $x$ data 
\cite{smclowx}
(not shown in Fig.~\ref{fig:g1world}) are correspondingly 
at a $Q^2$ well below 1 GeV$^2$, and the asymmetries are found 
to be compatible with zero.

With proton and neutron data available one can measure the 
Bjorken sum-rule $\int g_1^p dx  -\int g_1^n dx$.
For example,
SMC finds $\Gamma^p_1- \Gamma^n_1= 0.174^{+0.024}_{-0.012}$
at $Q^2 = 5$GeV$^2$
which is in excellent agreement with theory 
$\Gamma^p_1- \Gamma^n_1= 0.181\pm 0.003$. 
In general, the Bjorken sum-rule is confirmed to 10\% as shown in 
Fig.~\ref{fig:sumrule}.

Similar to the unpolarized data global NLO perturbative QCD analyses 
are performed
on the polarized structure function data sets. 
Fits are performed in different schemes, 
e.g. 
the AB and $\overline{\rm MS}$ schemes.
In the $\overline{\rm MS}$ scheme the polarized gluon distribution 
does not contribute explictly to the first moment of $g_1$. 
In the AB scheme on the other hand the axial anomaly 
$\alpha_s \Delta g$ does contribute explicitly to the first moment.
One finds for the  $\overline{\rm MS}$ (AB) scheme at a $Q^2$ of 
1 GeV$^2$ \cite{smcqcd}:
$\Delta \Sigma = 0.19\pm 0.05 (0.38\pm0.03)$ and 
$\Delta g = 0.25^{+0.29}_{-0.22} (1.0^{+1.2}_{-0.3})$ \cite{smcqcd}
where
$\Delta \Sigma = (\Delta u + \Delta d + \Delta s)$.
The main source of error in the QCD fits
comes from lack of knowledge about $g_1$ in the small $x$ 
region and (theoretical) 
the functional form chosen for the quark and gluon distributions in 
the fits \cite{abfr,def,smcqcd}.
Note that these QCD fits in both the AB and 
${\overline {\rm MS}}$ schemes give values of 
$\Delta \Sigma$ which are smaller than the Ellis-Jaffe value 0.6.

New fits are now being produced taking into account all the available 
data \cite{blum}.
At this workshop a new fit to the data was presented 
in~\cite{lichtenstadt}, 
which includes more recent $g_1$ data. 
Here one finds in the AB scheme and for Q$^2 = 1 $ GeV$^2$: 
$\Delta \Sigma= 0.40 \pm 0.02 (stat.)$ and 
$\Delta g = 0.63^{+0.20}_{-0.19} (stat.)$, 
i.e. a somewhat lower value for the polarized gluon contribution.
It is interesting to observe that this latest value 
for $\Delta g$ is in agreement with the prediction \cite{bbs} 
based on colour coherence and perturbative QCD.

In these pQCD analyses one ends up with a consistent picture of 
the proton spin:
the low value of $\Delta \Sigma$ may be compensated by a 
large polarized gluon. 
The precision on $\Delta g$ is however still rather modest. 
Moreover, it is vital to validate this model with {\it direct} 
measurements of $\Delta g$, as we discuss in the next section.
Also, the first moments depend on integrations from $x = 0$ to 1.
Perhaps there is an additional component at very small $x$?

Parton distributions with experimental and theoretical uncertainties,
resulting from the fits, are given in Fig.~\ref{fig:partons}.

Some progress on $g_1$ will come from
the improved precision with the new HERMES data and perhaps
remeasurements in the SMC kinematic range by COMPASS, but no 
extension of the kinematic range.
The kinematical reach is shown in
Fig.~\ref{fig:kin} for the different
machines discussed in the previous section, together with
the region covered by present fixed target experiments.
The region can be extended by several orders of magnitude 
both in $x$ and $Q^2$. 

A projection for a $g_1$ measurement at polarized HERA is shown in 
Fig.~\ref{fig:g1p}.
The low-$x$ behaviour of $g_1$ is indicated in the figure:
the straight line is an 
extrapolation based on Regge phenomenology, and the upper curve
presents a scenario suggested in~\cite{abhay} where 
$g_1$ rises as $1/(x \ln^2(x))$, which is the maximally singular 
behaviour still consistent with integrability requirements.
The low $x$ behaviour of $g_1$ by itself is an interesting 
topic as discussed in~\cite{kwiecinski}.
Small $x$ measurements \cite{badelek}, 
besides reducing the error on the first moment and $\Delta g$, would 
provide valuable information about perturbative QCD dynamics at low $x$,
where the shape of $g_1$ is particularly 
sensitive to the effects of
$(\alpha_s \ln^2 {1 \over x})^k$ resummation and DGLAP evolution
\cite{kwiecinski},
and
the non-perturbative QCD 
``confinement physics''
to hard (perturbative QCD) scale transition. 
Much larger changes in the effective intercept which describes the 
shape of the structure functions at small Bjorken $x$ are expected 
in $g_1$ than in the unpolarized structure function $F_2$ so far
studied at HERA as one increases $Q^2$ through the transition 
region from photoproduction to hard (deep inelastic) values of $Q^2$
\cite{sbadr}.
Polarized HERA would allow us to penetrate deeper into the small $x$
region; 
EIC would be especially suited to studies of the transition region.

An EIC has as expected a more restricted range, as shown in 
Fig.~\ref{fig:g1peic}. 
However the precision is substantially higher due to the 
high luminosity of 4 fb$^{-1}$/year.
THERA would reach $x$ values down to 10$^{-6}$, but needs at least 
one fb$^{-1}$ of data.

Note that the expected asymmetries at low $x$, i.e. $x \sim 10^{-4}$ are 
relatively small, about $10^{-3}$, which puts strong requirements
on the control of the systematic effects, as discussed in~\cite{works}.
At the lowest $x$ values at THERA,  the asymmetries
may be  even smaller, and the potential to actually 
extract a polarized structure function at these values still needs to 
demonstrated.

Looking in a little more detail to Fig.~\ref{fig:g1world} 
we see that the present data at small-$x$ correspond to a 
negative asymmetry and hence there must be a crossover at 
some intermediate $x$ value. 
Locating the crossover is an important experimental challenge. 
From the theoretical point of view the value of $x$ at 
which this occurs for the neutron spin asymmetry is the result 
of a competition between the SU(6) valence structure 
\cite{Close:1988br} and the chiral corrections 
\cite{Schreiber:1988uw,Steffens:1995at}.
Recent progress on extracting the neutron spin structure 
function from nuclear targets is reported in \cite{athomas}.

The large $x$ region (close to one) is also especially interesting and
is particularly sensitive to the valence structure of the nucleon.
In the spin dependent case we have no idea 
whatsoever of 
the behaviour of $g_{1n}$ beyond $x \sim 0.4$. 
On the basis of both perturbative QCD {\it and} SU(6) 
\cite{Close:1988br}, 
one expects the ratio of polarized to unpolarized structure functions,
$A_{1n}$,
should 
approach 1 as $x \rightarrow 1$ \cite{Melnitchouk:1996zg,Isgur:1999yb}. 
It is vital
to test this prediction. If it fails we understand nothing about the
valence spin structure of the nucleon! 
A precision measurement of $A_{1n}$ up to $x \sim 0.8$ 
will be possible following the 12 GeV upgrade of Jefferson Laboratory
(see Fig.~\ref{fig:jlab12}).

With proton and neutron data available one can measure the Bjorken 
sum-rule $\int g_1^p dx  -\int g_1^ndx$.
EIC data would allow one to measure this sum-rule, a key test for QCD, 
to the order of 1\% precision.

\section{THE POLARIZED GLUON DISTRIBUTION $\Delta g(x,Q^2)$}
\vspace{1mm}
\noindent

In the present understanding of the spin puzzle, measuring directly
the polarized gluon is the ``Holy Grail'' in spin physics.
The precision from QCD fits is still limited, and the ``Quest'' is on.

\begin{figure}[t!]
\begin{center}
\epsfig{file=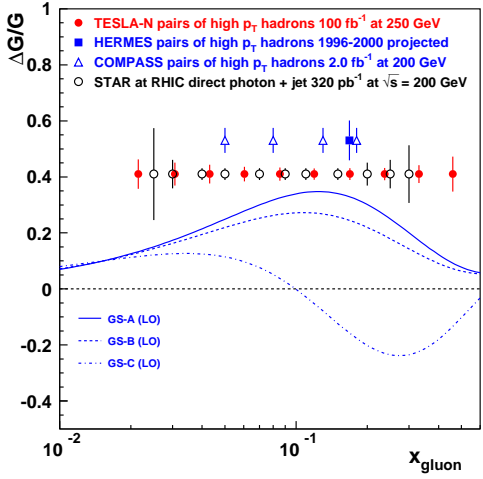,bbllx=0pt,bblly=0pt,bburx=570pt,bbury=550pt,width=7cm}
\caption{
Projected statistical accuracies for the measurement 
of $\Delta g/g(x)$ at TESLA-N, based on a integrated luminosity of 
100 fb$^{-1}$ in comparison with 
predictions from COMPASS, HERMES and RHIC. The phenomenological predictions
are calculated for $Q^2= 10 $ GeV$^2$\protect\cite{nowak}.}
\label{fig:deltag}
\end{center}
\end{figure}

COMPASS has been conceived to measure $\Delta g$ via the study of the
photon gluon fusion process, as shown in Fig.~\ref{fig:bgf}.
Hence the cross section of this process is directly related to the 
gluon density at the Born level. The experimental technique
consists in the reconstruction of charmed mesons.
The final uncertainty on the measurement of $\Delta g (x) /g (x)$
from open charm is estimated to be $\delta(\Delta g(x)/g(x))$ =0.11.
Additionally COMPASS will use 
the same process but with high $p_t$ particles instead of charm to 
access $\Delta g$. 
This may lead to samples with larger statistics, but these 
have larger background contributions, namely from QCD Compton processes
and fragmentation.
The expected sensitivity on the measurement of $\Delta g/g$
from high $p_t$ tracks is estimated to be $\delta(\Delta g/g) =0.05$,
and shown in 
Fig.~\ref{fig:deltag}.

\begin{figure}[t!]
\begin{center}
\epsfig{file=bgf.eps,bbllx=0pt,bblly=0pt,bburx=550pt,bbury=550pt,width=5cm}
\caption{$c\overline{c}$ production in Photon Gluon Fusion}
\label{fig:bgf}
\end{center}
\end{figure}

HERMES was the first to attempt to measure $\Delta g$, from high $p_t$ 
charged particles, as proposed for COMPASS above. 
The  measurement is at the limit of where the 
perturbative treatment of the data can be expected to be valid, but the 
result is interesting:
$\Delta g/g = 0.41 \pm 0.18 \pm 0.03$ at an average $<x_g>= 0.17$
\cite{fantoni}.
The expectation for the HERMES data collected until now is shown in
 Fig.~\ref{fig:deltag}.

TESLA-N could perform similar analyses as COMPASS and HERMES. 
Its high luminosity would allow for precision measurements in the 
range of (gluon) $x$: $0.02<x<0.4$, as shown in Fig.~\ref{fig:deltag}.

The hunt for $\Delta g$ is also one of the main physics drives for
polarized RHIC.  The key processes used here are high-$p_t$ prompt 
photon production 
$pp \rightarrow \gamma X$, 
jet production 
$pp \rightarrow $ jets $+ X$, 
and heavy flavour production
$pp \rightarrow c\overline{c}X, b\overline{b}X, J/\psi X$. 
Due to the first stage detector capabilities most emphasis 
has so far been put on the prompt photon channel. 
The expected precision is plotted in Fig.~\ref{fig:deltag}.
These anticipated RHIC measurements of $\Delta g$ have inspired 
new theoretical developments \cite{vogelsang} aimed 
at implementing higher-order calculations of 
partonic cross-sections into global 
analyses of polarized parton distribution functions, which will
benefit the analyses of future polarized $pp$ data to measure $\Delta g$.

Future $ep$ colliders can add information in two ways: by extending 
the kinematic range for measurements of $g_1$ or by direct 
measurements of $\Delta g$.
Including the
future polarized HERA $g_1$ data will improve the experimental error 
$\delta (\int \Delta g dx) $ to about 0.2.
The theoretical error 
is expected to decrease by more than a factor 2 once $g_1(x,Q^2)$ is 
measured at low $x$.
At an EIC one expects to reduce the statistical uncertainty even to 0.08
with about 10 fb$^{-1}$, as discussed
in ~\cite{lichtenstadt}.

It is however crucial for our full understanding of the proton spin that
the prediction of a large polarized gluon is confirmed or refuted 
by direct measurements
of $\Delta g$.
HERA has shown
that the large centre of mass system (CMS) energy allows
for several processes to be used to extract the unpolarized gluon 
distribution.
These include jet 
and high $p_t$ hadron production,  charm production 
both in DIS and photoproduction, and correlations between 
multiplicities of the current and target hemisphere of the events
in the Breit frame.


\begin{figure}[htb]
\epsfxsize=8.5cm
\epsffile[40 260 590 600]{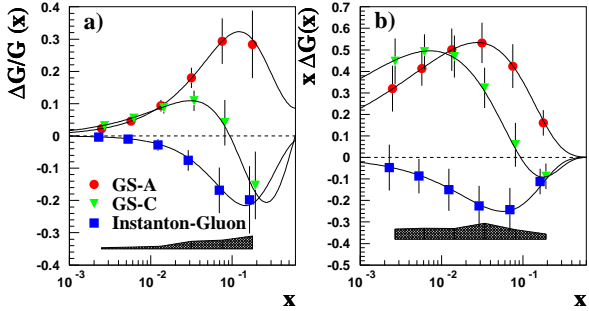}
\caption{Sensitivity to $\Delta g/g$ (a) and $x\Delta g$ (b) at HERA 
for a luminosity of $500~{\rm pb}^{-1}$ and three different assumptions
for the shape of $\Delta g(x)$. The error bars represent statistical
errors. The shaded band gives an estimate of the 
systematics \protect \cite{radel}. }
\label{fig:heralo}
\end{figure}

\begin{figure}[htb]
\epsfxsize=8.5cm
\epsffile[25 258 580 580]{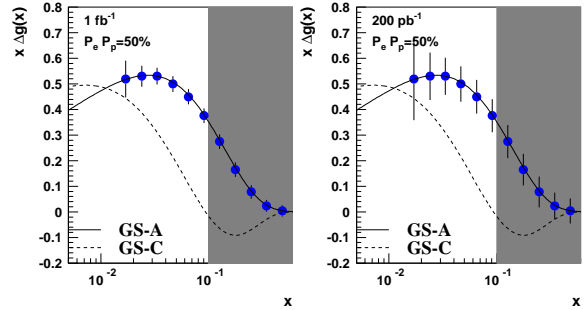}
\caption{The statistical precision of $x\Delta g$ from di-jets at LO
for EIC, for two different luminosities, with predictions of GS-A and
GS-C \protect \cite{radel}.}
\label{fig:glufig2}
\end{figure}

\begin{figure}[htb]
\epsfxsize=7.5cm 
\epsffile[40 160 550 680]{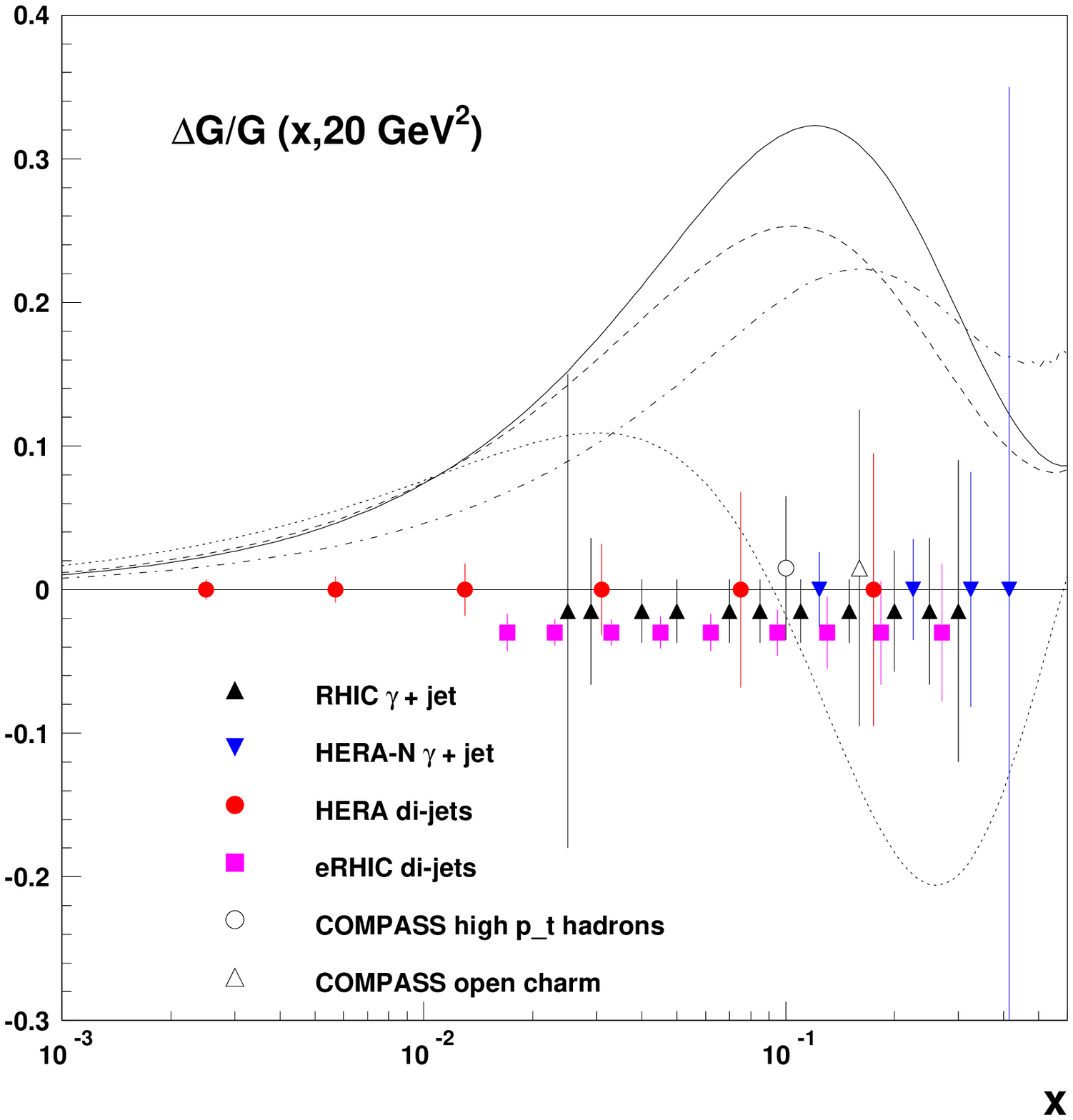}
\caption{Summary of the sensitivity of future measurements
of $\Delta g/g$ for HERA (500 fb$^{-1}$) and EIC/eRHIC (4 fb$^{-1}$)
compared with other  experiments (see text)\protect \cite{radel}.}
\label{fig:sumup}
\end{figure}

The most promising process for  a direct extraction of $\Delta g$ at 
HERA is  di-jet production, as discussed in ~\cite{radel}.
The underlying idea is to isolate boson-gluon fusion events, i.e.~a process
where the gluon distribution enters at the Born level. 
Jets are selected
with a $p_t>5$~GeV and are restricted to the acceptance 
of a typical existing HERA detector by 
the requirement $|\eta^{jet}_{LAB}| < 2.8$, where
$\eta^{jet}_{LAB}$ is the pseudo-rapidity in the laboratory system.
The resulting measurable range in  
$x$ (of the gluon) is $0.002 < x < 0.2$.
The results are shown in Fig.~\ref{fig:heralo} and compared to 
several predicted gluon distributions.
At EIC
the measurable range is reduced to  
$0.02 < x < 0.3$, as shown in Fig.~\ref{fig:glufig2}.
The reach at THERA is  $0.0005 < x < 0.1$ but an event sample of 
several hundred  pb$^{-1}$ will be needed.

These measurements allow for determination of the shape of $\Delta g(x)$ 
over a large region of $x$~\cite{nowak}.
The errors on the individual points for the di-jet measurement on 
$\Delta g(x)/g(x)$ are in the range from 0.007 to 0.1.
The total error on $\Delta g(x)/g(x)$ in the complete range is 0.02.

The global result is shown in Fig.~\ref{fig:sumup}, for HERA and 
EIC and compared to results from other planned or possible 
future polarized experiments: $pp$ scattering at RHIC~\cite{saito}
(STAR, $\sqrt{s}=200$ GeV),
$\mu p$ scattering in COMPASS~\cite{kunne}, and $pp$ scattering in
HERA-N~\cite{heran}.
All measurements are shown at LO. 
Clearly EIC can produce the (statistically) most precise 
measurements, and HERA covers the range to lowest possible $x$ values 
which can only be beaten by THERA, if it would be able to deliver 
sufficient luminosity.

An exploratory study was made \cite{abhay}, using the values of 
$\Delta g(x)$
obtained from the di-jet analysis as an extra constraint in the fit of 
$g_1$ data discussed above.
The improvement of the errors on the first moment of $\Delta g$ due to 
the inclusion of di-jet data is shown in Table~\ref{tab-hera}.
\begin{table}[t]
\caption{\label{tab-hera} The expected statistical 
uncertainty in the determination of
  the first moment of the gluon distribution at $Q^{2} = 1$ GeV$^2$
(500~pb$^{-1}$), see text.}
\hfil
\begin{tabular}{||l|c||}
\hline\hline
  {\bf Analysis Type} &  {\boldmath $\delta (\int \Delta g~{\rm d}x)$} \\
\hline \hline
1. $g_1$ fixed target              &   0.3           \\
\hline
2.  $g_1$ fixed target + HERA   &   0.2           \\
\hline
3. di-jets at HERA                   &   0.2          \\
\hline
4. combined 2 \& 3                    &    0.1   \\
\hline\hline
\end{tabular}
\hfil
\end{table}
The first two rows give the values quoted before, namely for the NLO QCD
analysis without and with projected
polarized HERA data for $g_1$. 
The third row shows the expected error if only the di-jet asymmetry is 
added to the fixed target $g_1$ data, and the fourth row shows the total 
improvement using all available information.
Hence the first moment of the gluon can be determined with a precision
of about 10\%.
Looking further to the future the projected error from one year of running 
with TESLA-N 
could be reduced to
$\delta (\int \Delta g~{\rm d}x) \pm 0.06$ \cite{teslan}.

\begin{figure}[htb]
\epsfxsize=7.5cm 
\epsffile[0 0 550 550]{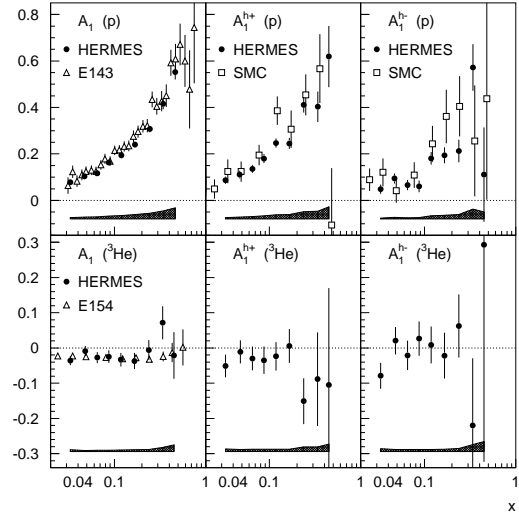}
\caption{Inclusive $A_1$ and semi-inclusive $A^h_1$ asymmetries form
polarized H and $^3$He targets. The data points are given at the measured
mean at each value of $x$, which is different for each 
experiment \protect \cite{fantoni}.}
\label{fig:semi}
\end{figure}

\begin{figure}[htb]
\epsfxsize=6.5cm 
\epsffile[80 80 500 730]{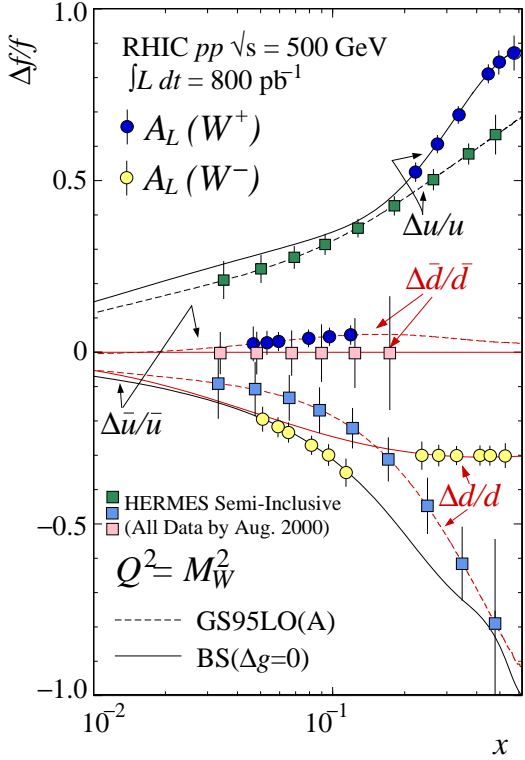}
\caption{Polarization of $u, d, \overline{u}, \overline{d}$ as 
functions of $x$ modelled by~\cite{bourely,GS}. 
Both the sensitivities of HERMES 
semi-inclusive DIS and RHIC  $W$ measurements are 
shown \protect \cite{saito}.}
\label{fig:rhicpdf}
\end{figure}

\begin{figure}[t!]
\epsfxsize=7.5cm 
\epsffile[0 0 580 830]{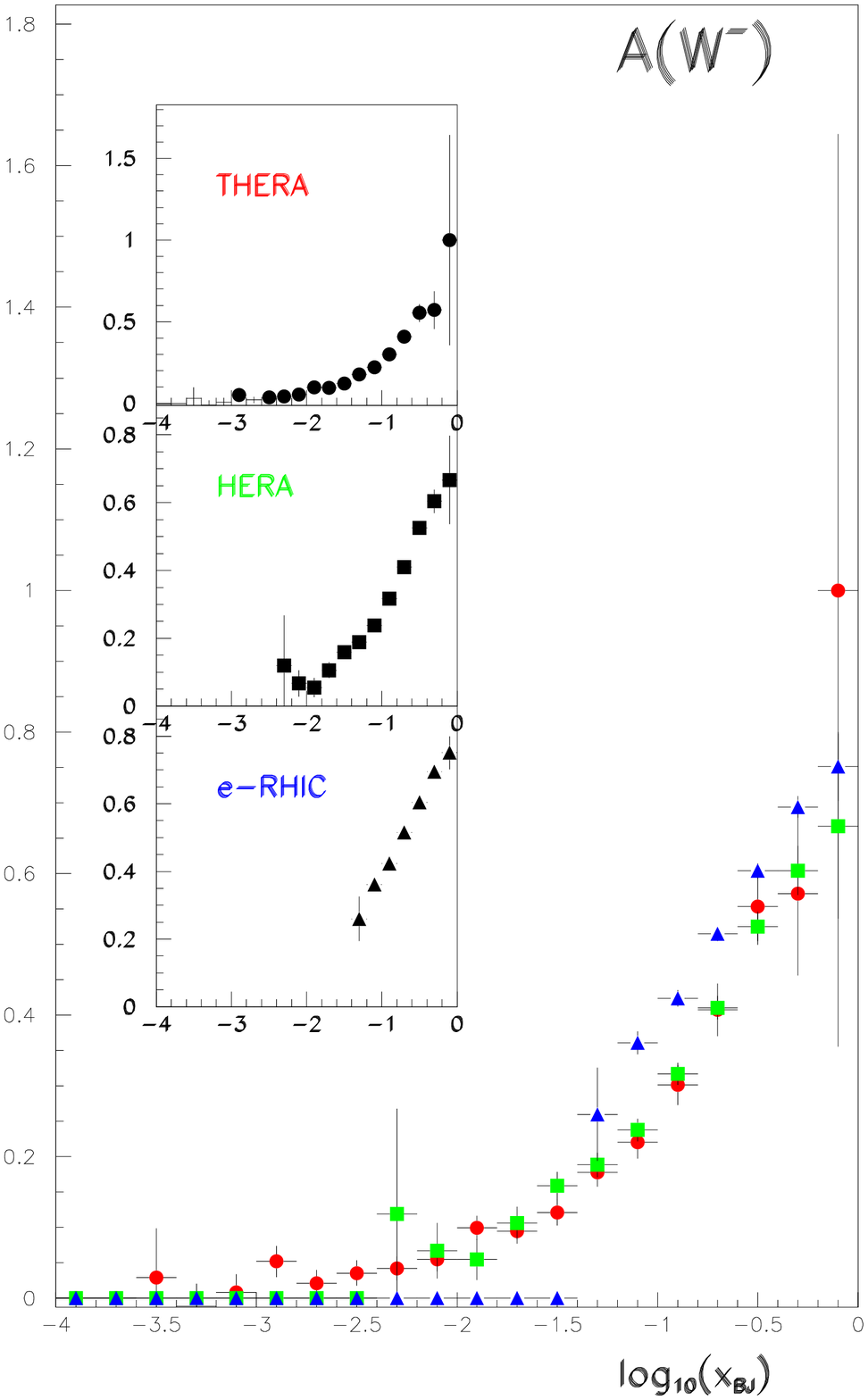}
\caption{Spin asymmetries $A^{W^-}$ 
   for charged current 
events for a total
  luminosity of 500 pb$^{-1}$. 
The error bars represent
  the statistical uncertainty of the 
measurement \protect \cite{contreras}.} 
\label{fig_results}
\end{figure}

\section{POLARIZED QUARK DISTRIBUTIONS}
\label{sec:pquark}
\vspace{1mm}
\noindent

In present fixed target experiments information on the flavour decomposition 
can be obtained from  semi-inclusive measurements, 
i.e.~measurements where a final state hadron is tagged,
in the current fragmentation region. 
The extracted spin distributions for the up, down and sea quarks, 
as obtained by the HERMES and SMC
collaborations are shown in Fig.~\ref{fig:semi}~\cite{fantoni}. 
For this extraction it was assumed that the polarization of the sea 
quarks is flavor independent.
Furthermore, the analysis was carried out using leading-order QCD
(the naive parton model).
The polarization of the up and down quarks are positive and negative
respectively, while the sea quark polarization remains negative within
the measured range and consistent with zero. 
In future the data of the RICH detector in the HERMES spectrometer, 
together with the large sample of deuterium data still to be analysed, 
will allow for more precise results with less assumptions.
Further accuracy will come from a next-to-leading order QCD analysis from
the data.

The number for $\Delta u - \Delta d$
extracted from these measurements is
$\Delta u - \Delta d = 0.84 \pm 0.07 \pm 0.06$ at $Q^2=2.5$GeV$^2$, 
which compares with the prediction $1.01 \pm 0.05$ for the Bjorken 
sum-rule at this $Q^2$ {\it after} higher-order radiative corrections 
have been applied in the Bjorken sum-rule.
While this comparison looks at first quite attractive it should be noted
that the data analysis is done using only leading order 
formulae, so that the correct quantity 
to compare with for full self-consistency is the value of 
$g_A^{(3)}$ ($=1.267$).
A next-to-leading order analysis of the data along the lines presented
in \cite{def} would be very useful.

Semi-inclusive measurements of $g_1$ in the target fragmentation region 
with a polarized $ep$ collider
would enable one to test the target 
\mbox{(in-)}dependence of the small value of $g_A^{(0)}|_{\rm pDIS}$
\cite{shoredef}.

The dependence on the details of the fragmentation process limits the 
accuracy of the method above. At RHIC~\cite{saito}
 however the polarization of the 
$u, \overline{u}, d$ and $\overline{d}$ quarks in the proton will 
be measured directly and precisely using $W$ boson production 
in $u\overline{d} \rightarrow W^+$ and $d\overline{u} \rightarrow
W^-$. 

The charged weak boson is produced through a pure V-A coupling
and the chirality of the quark and anti-quark in the reaction is fixed. 
A parity violating asymmetry for $W^+$ production in $pp$ 
collisions can be expressed as
\begin{equation}
 A(W^+) = 
\frac{\Delta u(x_1)\overline{d}(x_2) -\Delta\overline{d}(x_1) u(x_2)}
{ u(x_1)\overline{d}(x_2) + \overline{d}(x_1) u(x_2)}
\end{equation}

For $W^-$ production $u$ and $d$ quarks should be exchanged. 
The expression converges to 
$\Delta u(x)/u(x)$ and $- \Delta\overline{d}(x)/\overline{d}(x)$ 
in the limits
$x_1 >> x_2$ and $x_2 >> x_1$ respectively. 
The momentum fractions are calculated as 
$x_1= \frac{M_W}{\sqrt{s}} e^{y_W}$ and 
$x_2= \frac{M_W}{\sqrt{s}} e^{-y_W}$, with $y_W$ the rapidity of 
the $W$. 
The experimental difficulty is that the $W$ is observed through its 
leptonic decay $W \rightarrow l \nu$ and only the charged lepton is 
observed.
With the assumed integrated luminosity of 800 pb$^{-1}$ at $\sqrt{s}
= 500 $ GeV, one can expect about 5000 events each for $W^+$ and $W^-$.
The resulting measurement precision is shown in Fig.~\ref{fig:rhicpdf}.

Furthermore, information on the quark densities will be extracted 
using Drell-Yan production into lepton pairs, and the measurement 
of inclusive particle production.

At HERA, apart from semi-inclusive 
studies,  one also has the option to study quark flavours  
via charged current interactions~\cite{deroeck,contreras}.
 The asymmetry is defined by 
\begin{equation}
A^{W\mp} =
\frac{d\sigma^{W^\mp}_{\uparrow\downarrow}-d\sigma^{W^\mp}_{\uparrow\uparrow}}
{d\sigma^{W^\mp}_{\uparrow\downarrow}+d\sigma^{W^\mp}_{\uparrow\uparrow}}
\nonumber\\
\approx \frac{g^{W^{\mp}}_5}{F^{W^{\mp}}_1} \label{eq_as}
\end{equation}
with 
$g^{W^-}_5 = \Delta u+\Delta c - \Delta\bar{d} - 
\Delta
\bar{s} $, $g^{W^+}_5 = \Delta d+\Delta s - \Delta\bar{u} - \Delta\bar{c}$.
~\cite{contreras}.
The total missing transverse momentum (which is a signal for the 
escaping neutrino) was required to be $P_{Tmiss}>15$ GeV,  and the region 
$Q^2>225$ GeV$^2$ has been selected for this analysis. 
The results for the asymmetries for both HERA, EIC and THERA
 are shown in Fig.~\ref{fig_results}. 
The error bars indicate the statistical precision 
of the measurement. The asymmetries are very large
and significant measurements can be produced at THERA down to 
$x= 10^{-3}$.
These charged current measurements, with both 
$e^+$ and $e^-$ beams,   can be used to 
extract e.g.~the $\Delta u$ and $\Delta d$ distributions.

It has been pointed out that neutrino factories are an ideal tool
for polarized quark flavour decomposition studies.
These would allow one to collect large data samples of charged current 
events, in a kinematic region $(Q^2,x)$ of the present fixed target 
data. 
A complete separation of all four flavours and anti-flavours 
becomes possible, and in particular the role of $\Delta s$ 
-- which may be quite different in 
e.g. `anomaly', `instanton' and `skyrmion' models -- can be determined.

Pseudo-data for charged current structure functions have been 
calculated~\cite{ridolfi} according to these three models,
and used in NLO pQCD fits. It was shown that the singlet quark moment 
can be improved  by a factor 2-3 or so. The Bjorken sum rule 
could be tested to the few \% level. C-even and C-odd 
polarized quark distributions can be measured with an accuracy of a few
\% for the $u$ and $d$ quark, while the strange components can be 
measured to the level of 10\%, which is sufficient to distinguish between
the models above.

Finally, we note that complementary measurements of the spin dependent 
parton distributions from different experimental conditions 
(in polarized $ep$ and $pp$ collisions) are necessary 
to experimentally check factorization in spin dependent processes --
that is, that the extracted parton distributions are indeed process
independent.
Factorization is commonly assumed to hold 
(beyond perturbative QCD) but has so far not been checked in experiment
for spin dependent processes.

\section{PHOTOPRODUCTION}
\vspace{1mm}
\noindent

Important information about QCD spin physics will also come from
polarized photoproduction measurements.
The fundamental Gerasimov-Drell-Hearn sum-rule (5) 
\cite{gerasimov}
($\int_{\nu_{th}}^{\infty} {d \nu \over \nu} (\sigma_A - \sigma_P)(\nu)
=
- {2 \pi^2 \alpha \kappa^2 \over m^2} 
$)
follows from general principles of quantum field theory: 
causality, unitarity, Lorentz and electromagnetic gauge invariance
plus one assumption: that we can use an unsubtracted dispersion 
relation for the spin dependent part of the forward Compton amplitude.
Modulo the no-subtraction hypothesis, 
the Gerasimov-Drell-Hearn sum-rule is valid 
for a target of arbitrary spin $S$, whether elementary or composite 
\cite{brod69}.
Regge arguments imply that the GDH integral is expected to converge.
Failure of the GDH sum-rule would imply a 
(finite) 
subtraction constant in the dispersion relation.
Such a term would follow from a $J=1$ 
Regge fixed pole with non-polynomial residue \cite{fixedpole}.
There is no such fixed pole in perturbation theory, 
at least up to
$O(\alpha^3)$ in QED.
Any violation of the GDH sum-rule would be most interesting indeed and
yield new challenges for theorists to explain the (non-)perturbative QCD 
dynamics of such an effect.

The ``generalized'' GDH integral
\begin{equation}
I (Q^2) =
{2 m^2 \over Q^2}
\int_0^1 dx g_1 (x, Q^2)
\end{equation}
(with
$I(0) = - {1 \over 4} \kappa^2 $)
changes sign between photoproduction and deep inelastic $Q^2$ 
(assuming that the GDH sum-rule is correct); 
the zero crossing point in $I(Q^2)$ is particularly sensitive to spin
structure in the resonance region \cite{ioffe}.
Dedicated low $Q^2$ investigations of $g_1$ are underway at Jefferson
Laboratory \cite{meziani}.

\begin{figure}[t!]
\begin{center}
~ \epsfig{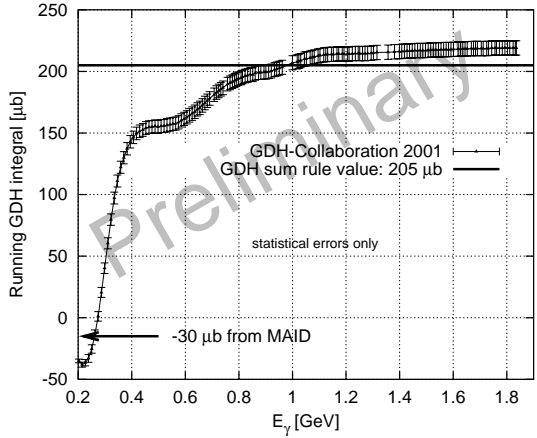}
\caption{Running GDH integral for the proton (ELSA and MAMI)\protect 
\cite{helbing}.}
\label{fig:gdhdata}
\end{center}
\end{figure}

Present experiments at ELSA and MAMI are aimed at measuring the GDH 
integrand through the range of incident photon energies 
$E_{\gamma} = 3.1$
0.14 - 0.8 GeV (MAMI) and 0.7 - 3.1 GeV (ELSA) \cite{helbing}.
The presently analysed GDH integral on the proton 
(up to 1.9 GeV) is shown in Fig.~\ref{fig:gdhdata}  and 
is dominated by the $\Delta$ resonance contribution.
Intriguing is that the preliminary GDH integral for the proton up to 
the presently analysed 1.9 GeV seems to be converging to a value two 
standard deviations greater than the GDH prediction.
Further experiments are planned to measure the neutron GDH integral
via a polarized deuteron target.

Clearly, it will be very important to measure the high-energy part 
of the sum-rule, both to check the sum-rule itself and also to test
spin dependent Regge theory.
First measurements will be made at SLAC for $E_{\gamma}$ between 5 and 
40 GeV for both proton and neutron.
Finally, at collider energies, it can be pointed out that a measurement 
of the total polarized photoproduction cross section 
$\Delta \sigma_{\gamma p} (\nu)$ as a function of the photon-proton CMS 
energy $\nu$ at HERA or EIC would contribute to a precise understanding 
of the Gerasimov-Drell-Hearn sum rule, as discussed in~\cite{sbadr}. 
In particular the  Regge behaviour of the energy dependence of the 
polarized cross section can be tested. 
Spin dependent Regge theory (totally untested at the present time) 
provides a baseline for investigations of the small $x$ behaviour of $g_1$ 
at deep inelastic values of $Q^2$.
The sensitivity which could be reached with EIC is about one order of 
magnitude larger than the one for HERA.

High-energy photoproduction processes have also been shown to be 
sensitive probes of the polarized parton structure in the proton 
{\it and} in the photon. 
These processes would allow us to obtain, 
for the first time,
unique information on the polarized structure of the photon.
Single jet production has been studied in~\cite{gehr}.
In~\cite{gamma} it was shown that  also a high $p_t$ track analyses 
yields a similar sensitivity, with only modest luminosity requirements.

\begin{figure}[t!]
\begin{center}
~ \epsfig{file=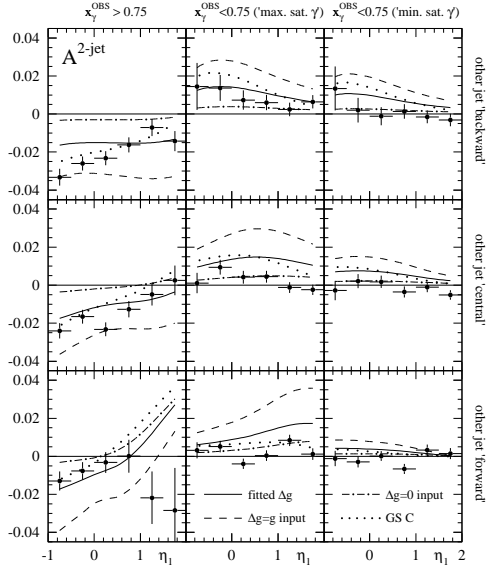,bbllx=0pt,bblly=0pt,
bburx=600pt,bbury=580pt,width=8cm}
%
\caption{Polarized photoproduction at HERA of di-jets ($E_{T,1} > 8$~GeV,
$E_{T,2} > 6$~GeV): 
asymmetries for direct (first column) and resolved (second and third 
column) photon contributions as function of the rapidity of the first 
jet and for different orientations of the second jet.
Second and third column correspond to different scenarios for 
the parton content of the polarized photon suggested 
in~\protect\cite{grvgamma}.
The error bars shown 
correspond to a Monte Carlo sample of 50~pb\protect{$^{-1}$} 
\protect \cite{gamma}.
}
\label{fig:phoprod}
\end{center}
\end{figure}

Polarized photoproduction of di-jets has been investigated in detail,
including in particular effects due to 
parton showering, hadronization, jet finding and jet clustering. 
It could be demonstrated that, although 
these effects yield sizable corrections, the measurable asymmetry will
be largely preserved at the hadron level~\cite{gamma}. 
An example for the correspondence of parton and hadron level 
asymmetries is 
shown in Fig.~\ref{fig:phoprod}, obtained with a moderate integrated 
luminosity of only 50~pb$^{-1}$. 
A first idea on the discriminative power on the photon 
structure of future measurements 
can however be gained by comparing the predictions obtained 
with the two (minimal and maximal) polarization scenarios 
proposed in~\cite{grvgamma}, as done in Fig.~\ref{fig:phoprod}.

\section{TRANSVERSITY}

A degree of freedom which is currently of considerable interest 
\cite{matthias,boer,anselmino}
is the density of transversely polarized quarks inside a transversely 
polarized proton.

The transversity can be interpreted in parton language as follows:
consider a nucleon moving with (infinite) momentum in the $\hat
e_{3}$-direction, but polarized along one of the directions transverse
to $\hat e_{3}$.  $\delta q_{a}(x,Q^{2})$ counts the quarks of flavor
$a$ and  momentum fraction $x$ with their spin parallel the spin of a
nucleon minus the number antiparallel.  If quarks moved
nonrelativistically in the nucleon, $\delta q$ and $\Delta q$ would
be identical, since rotations and Euclidean boosts commute and a
series of boosts and rotations can convert a longitudinally polarized
nucleon into a transversely polarized nucleon at infinite momentum. 
So the difference between the transversity and helicity distributions
reflects the relativistic character of quark motion in the nucleon. 
There are other important differences between transversity and
helicity.  For example, quark and gluon helicity distributions
($\Delta q$ and $\Delta g$) mix under $Q^{2}$-evolution.  There is no
analog of gluon transversity in the nucleon, so $\delta q$ evolves
without mixing, like a nonsinglet distribution function.  The lowest
moment of the transversity is proportional to the nucleon matrix
element of the tensor charge, $\bar q i\sigma^{0i}\gamma_{5}q$, which
couples only to valence quarks (it is $C$-odd). 
Not coupling to glue or $\bar q q$ pairs, the tensor charge promises 
to be more quark-model--like than the axial charge and should be an 
interesting contrast \cite{jaffe01}.

The experimental study of transversity distributions at leading
twist requires observables which are the product of two 
objects with odd chirality.
In $pp$ collisions the transverse double spin asymmetry, 
$A_{TT}$, is proportional to $\delta q \delta\bar{q}$ 
with even chirality. 
Recent studies show however that these could only be accessed 
at RHIC for very large luminosity samples, perhaps after a 
possible upgrade.

The HERMES experiment at DESY is expected to start measurements with
transverse target polarization in the near future.
The observable is Collins' single spin asymmetry 
\cite{collins}
for 
charged pion production in deep inelastic electron proton scattering
The azimuthal distribution of the final state pions with respect to the virtual
photon axis carries information about the transverse quark spin orientation.
In a partonic picture of
the nucleon the transverse single spin asymmetry $A_T$ is related to
the transversity distributions as follows:
\begin{equation}
A_T(x,z) \sim \frac{\sum_q e^2_q\delta q(x)H_1^{\perp q}(z)}
                    {\sum_q e^2_q  q(x)D_1^{q}(z)}
\end{equation}
where $H_1^{\perp q} $ is the Collins function for a quark of flavor $q$ and $D_1^q$ 
is the regular spin independent fragmentation function.
An analysis
procedure for the HERMES transverse asymmetries has been 
proposed~\cite{korotkov}
that
results in the extraction of the shape for $\delta u(x)$ and 
the ratio 
$H_1^{\perp}/D_1^u$.
It is
assumed that the $u$-quark dominates and
$\delta u(x_0=0.25)=\Delta u(x_0)$ at some soft scale, 
$Q_0^2 \approx 0.4$\, GeV. Alternatively, one could try to measure
the ratio $H_1^{\perp}/D_1^u$ from fragmentation function
measurements in $e^+e^-$ annihilation using LEP and B-factory data. 

\begin{figure}[htb]
\vskip -0.2cm
~ \epsfig{file=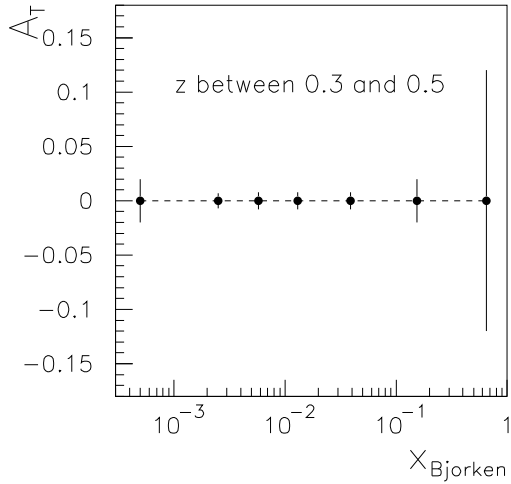,bbllx=0pt,bblly=400pt,
bburx=300pt,bbury=700pt,width=8cm}
\vskip -0.7cm
\caption{Projected sensitivities for the single spin asymmetries in single
inclusive pion production at the EIC for $\int_{day} Ldt=20$\,pb$^{-1}$ and
$0.3<z=E_{\pi}/E_{\gamma}<0.5$\protect \cite{matthias}.}
\label{fig:eic}
\end{figure}

HERMES will measure   
$\delta u(x)$ between $0.02<x<0.7$ with a statistical 
error of about $\delta u \pm 0.3$ at the lowest $x$ and $\delta u \pm 0.1$ at
highest $x$ \cite{nowak} and 
comparable systematic errors.
COMPASS also plans such a measurement.
The higher beam energy, $100-200$\,GeV compared to $27.5$\,GeV 
may allow, 
for the
first time, to explore the  interference fragmentation processes. 

However,
the
most promising experiments to measure transversity distributions 
and their first moments,
the tensor charge are EIC and TESLA-N.
For TESLA-N 
the first moment of $\delta u$ can be measured to about one percent
and for $\delta d$ to $6$\% \cite{nowak}. 
A measurement with the statistical precision of HERMES can be reached 
at the EIC, but at lower $x$, see Fig.~\ref{fig:eic} for only 20
pb$^{-1}$.

For polarized hadron colliders such as RHIC, 
in addition to the double spin
asymmetry
measurements in Drell Yan, several 
single spin asymmetry measurements have been suggested:
single spin asymmetries in two pion interference fragmentation;
 the measurement of azimuthal asymmetries in the production of
      pions around the jet-axis (Collins fragmentation function);
 interference fragmentation and Collins fragmentation in direct photon 
      jet events.
With the presently available luminosities and detectors
the proposal of Collins et al. \cite{collins2}
and Jaffe et al.~\cite{jjt} to
utilize chiral odd two pion interference fragmentation processes
appears to be the most promising approach to measure transversity
at RHIC.

All future transversity measurements have in common
that their success critically depends on the knowledge of the spin dependent
and chiral odd fragmentation functions giving sensitivity to transverse quark
spin in the final state. A detailed recipe
how to extract interference fragmentation functions  
from light-quark di-jet events in $e^+e^-$ collisions 
is given in~\cite{artru69}. 
A recent discussion of the prospects of transversity measurements using 
future fragmentation function information from $e^+e^-$ can be found in 
reference~\cite{danieldis}.

Additional studies in $e^+e^-$ have been proposed using the high statistics 
data sample of the BELLE experiment at the KEK B-factory \cite{intent}
with the goal to measure the two relevant fragmentation functions:
namely the Collins function $H_1^{\perp}$ and 
the interference fragmentation functions $\delta \hat{q}^{h_1,h_2}$:
For the first one measures
the fragmentation of a transversely polarized quark into a charged pion
and the azimuthal distribution of the final state pion with respect to
the initial quark momentum (jet-axis).
For the second one measures the fragmentation of transversely polarized
quarks into
pairs of hadrons in a state which is the superposition of two different
partial wave amplitudes; e.g. $\pi^+,\pi^-$ pairs in the $\rho,\sigma$ 
invariant mass region.
The high luminosity and particle identification capabilities of detectors
at B-factories makes these measurements possible.

In all transversity measurements have a bright future and may reveal some 
surprises.

\section{DVCS AND EXCLUSIVE PROCESSES}

Deeply virtual Compton scattering (DVCS) provides a possible experimental 
tool to access the quark total angular momentum, $J_q$, in the proton 
through generalized parton distributions~\cite{ji}.
The form-factors which appear in the forward limit $(t \rightarrow 0)$ 
of the second moment of the spin-independent generalized quark parton 
distribution in the (leading-twist) spin-independent part of the DVCS 
amplitude project out the quark total angular momentum defined through 
the proton matrix element of the QCD angular-momentum tensor. 
Going from $J_q$ to the orbital angular momentum $L_q$ is non-trivial 
in that one has to be careful to quote the perturbative QCD factorization 
scheme and process 
(polarized deep inelastic or $\nu p$ elastic scattering) 
used to extract information about the intrinsic spin contribution 
$S_q$ \cite{sbdvcs}.
\footnote{
The quark total angular momentum $J_q$ is anomaly free in QCD so that 
QCD axial anomaly effects occur with equal magnitude and opposite sign
in $L_q$ and $S_q$.
$L_q$ is measured by the proton matrix element of 
$[{\bar q} ({\vec z} \ {\rm x} \ {\vec D})_3 q](0)$.
Besides its sensitivity to the axial anomaly, 
$L^q$ is also sensitive to gluonic degrees of 
freedom through the 
gauge-covariant derivative --- for a recent discussion see~\cite{jaffe01}.}

DVCS studies have to be careful to chose the kinematics not to be 
saturated
by a large Bethe-Heitler background where the emitted real photon 
is radiated from the electron rather than the proton.
The HERMES and Jefferson Laboratory experiments
measure in the kinematics where they expect to be dominated by 
the DVCS-BH
interference term and observe the $\sin \phi$ 
azimuthal angle and 
helicity dependence expected for this contribution.
A first measurement of the single spin asymmetry 
has been 
discussed~\cite{amarian,hermesdvcs}, 
which has the characteristics expected from the 
DVCS-BH interference.
Recent JLab measurements are reported in~\cite{clas}.

Further measurements have been performed on exclusive $\pi^+$ production.
At lower energies virtual Compton scattering experiments at 
Jefferson Laboratory and MAMI probe the electromagnetic deformation of 
the proton and measure the electric and magnetic polarizabilities of the 
target. Preliminary cross section measurements show clear effects of 
polarizabilities~\cite{disalvo}.

Higher twist corrections to the helicity-conserving DVCS amplitude 
appear 
not to be dramatically large. 
In contrast, 
the twist-two photon helicity-flip amplitude in DVCS 
starts at order $\alpha_s$.
Here leading higher-twist correction may arise at 
the tree-level already and 
therefore be large as compared to the leading-twist amplitude~\cite{lech}.

\begin{figure}[t!]
\begin{center}
\epsfig{file=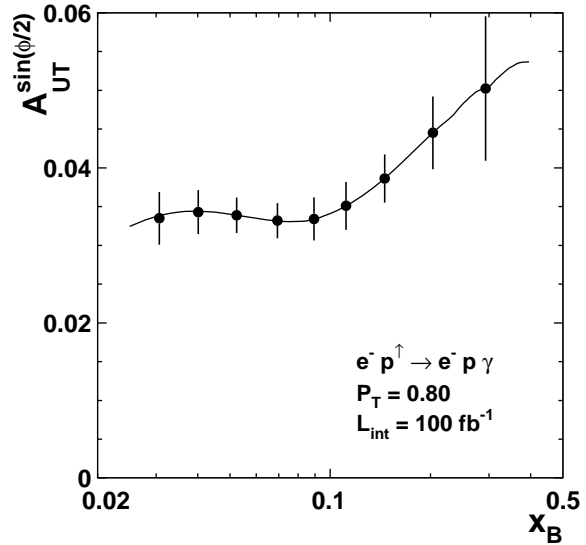,bbllx=0pt,bblly=0pt,bburx=570pt,bbury=570pt,width=8cm}
\caption{Projected statistical accuracy for the $x_B$ dependence of the 
$\cos(\phi/2)$ moment in a high-luminosity experiment
with unpolarized electron beam and polarized target, 
including detector acceptance \protect \cite{nowak}. }
\label{fig:dvcs}
\end{center}
\end{figure}

In future the HERMES and COMPASS experiments will explore DVCS measurements 
further. 
HERMES plans to run in 2004-2006 with an unpolarized target~\cite{nowak}, 
allowing one to collect large statistics samples.
This will allow one to study the $t$ dependence of these asymmetries, and 
perhaps even a first approximate determination of the GPD $H$, 
for the $u$ quark. 
Since also the knowledge on the GPD $E$ is necessary
to make contact with $J_q$, precision data with an unpolarized beam 
on a transversely polarized target would be needed. Even with 3 years of 
data with a (factor 2) improved target density, it seems marginal to achieve 
that goal.

A real breakthrough could be achieved with a  high statistics
experiment such as TESLA-N~\cite{nowak}. Fig.~\ref{fig:dvcs} shows a typical observable
at such an experiment: the $\cos (\phi/2)$ moment of the cross section
asymmetry for a 30 GeV beam on a $NH_3$ target. The error bars
are for one year of running. It shows that even the
measurement of two-dimensional dependences ($x$ and $t$) is within reach.
It is therefore hopeful that the GPD's can be mapped in future, and 
perhaps deeper insight into the angular momentum structure can be gained.

In other exclusive channels, 
measurements of exclusive coherent vector meson production have been 
proposed~\cite{moinester} to look for colour transparency effects at 
COMPASS.
Measurements of low-energy form-factors in exclusive strangeness 
production processes have been used to test models of confining 
constituent quark propagators \cite{fischer}.
The spin structure of light-cone wavefunctions \cite{hwang} may be probed
in hard exclusive processes.

\section{POLARIZATION AND NEW PHYSICS}
\vspace{1mm}
\noindent

\begin{figure}[htb]
\epsfxsize=7.5cm 
\epsffile[50 50 540 780]{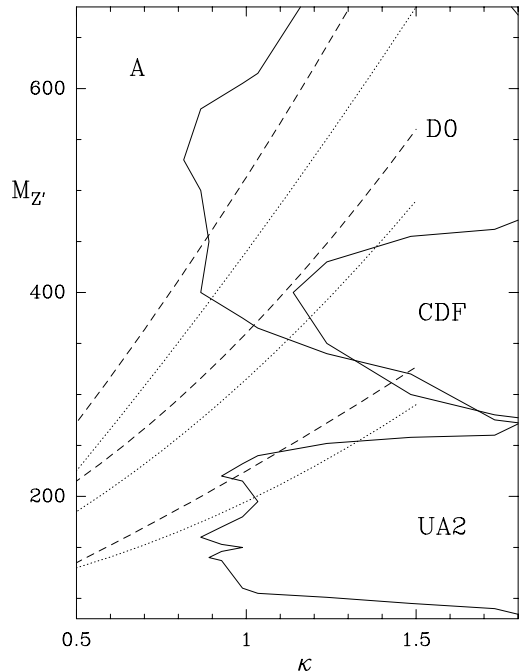}
\caption{Bounds on the parameter space for leptophobic flipped SU(5)
models,
see \protect ~\cite{virey}.}
\label{fig:rhicbsm}
\end{figure}

\begin{figure}[htb]
\epsfxsize=7.5cm 
\epsffile[50 50 540 780]{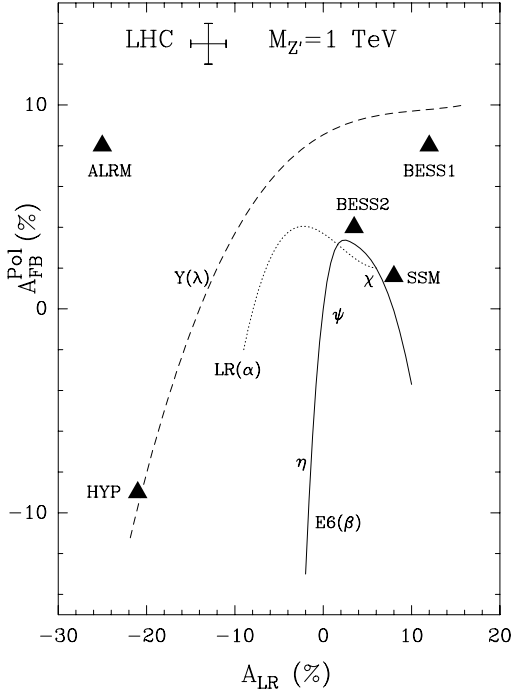}
\caption{$A^{\rm pol}_{FB}$ versus $A_{LR}$ according to the various models,
see \protect ~\cite{virey}.}
\label{fig:lhcbsm}
\end{figure}

The recent precision measurement of the muon's anomalous magnetic moment 
$(g-2)$ at BNL~\cite{hughes}
has revealed a 2.6 standard deviations difference from the Standard Model 
prediction, and created renewed excitement about possible ``new physics''.

The use of high energy polarized beams for discovering and 
in particular disentangling properties of new physics 
beyond the Standard Model, has been studied in ~\cite{sofferssc,rhicvirey}.
Polarized beams allow one to define new observables: the spin asymmetries.
These asymmetries typically involve ratios of cross sections, 
thus minimizing the systematics which often dominate
the precision, a well known example 
being the  high $E_t$ production at the Tevatron. Hence 
such asymmetries can improve the discovery potential of an accelerator.
Moreover such spin asymmetries offer the opportunity
to extract information on the chiral structure of the new interaction.

For RHIC, some specific examples have been studied and 
reported~\cite{lhcvirey}. E.g. a useful variable is
the single parity violating
asymmetry 
\begin{equation}
A_L = \frac{d\sigma(-)-d\sigma(+)}{d\sigma(-)+d\sigma(+)}
\end{equation}
for single jet production cross sections ($ pp \rightarrow jet +X$)
where only one of the protons is polarized.
A contact interaction analysis of this variable 
yields the limits as given in 
Table~\ref{tab:rhic}.

\begin{table}
\begin{tabular}{ccccc}
\hline
$L({\rm fb}^{-1})$ &  0.8 & 4 & 20 & 100\\

$\Lambda$ (TeV) &  3.2 & 4.55 & 6.15 & 7.55\\

\hline
\end{tabular}
\caption{Limits on $\Lambda_{LL}$ at 95\% for RHIC with $\sqrt{s}$
= 500 GeV}
\label{tab:rhic}
\end{table}

At the Tevatron the expected sensitivities to $\Lambda$
(the compositeness scale)
are 3.2 TeV (3.7 TeV) for a 1 fb$^{-1}$ (10 fb$^{-1}$)
data sample~\cite{lhcvirey}. Hence for comparable luminosity 
the polarized RHIC sensitivity is considerably larger, despite
the lower energy.

Another example is  leptophobic $Z'$ production, i.e. new $Z' $
gauge bosons which couple very weakly or not at all
to leptons. Fig.~\ref{fig:rhicbsm}
shows the constraints on the parameter space $\kappa = g_{Z'}/g_Z$ and
the mass $M_{Z'}$. Dashed curves are for RHIC corresponding
 to $\sqrt{s}=650$ GeV and dotted to $\sqrt{s}=500$ GeV.
From bottom to top the curves  correspond to $L= 1, 10$ and 100 fb$^{-1}$.
The increase in luminosity 
 seems to be more effective than the increase
in energy.
In conclusion one finds that RHIC is competitive/complementary with the
 Tevatron to discover a new weak force belonging purely to the quark
sector.

The LHC will give access to a complete new energy domain, and 
is being built with the mission to discover new physics. 
Once found the main task will 
be to try to understand it and disentangle the underlying dynamics.
Generally, one finds that if the signals of new physics are leptonic,
 then polarization will not be needed to 
discover it. 
However the following example shows the power to interpret the results.
Assume that a new $Z'$ boson is observed:
$pp \rightarrow l^+l^-X$. With one polarized beam
one can define 2 parity violating asymmetries:
\begin{eqnarray}
A_{LR} &=& \frac{d\sigma^- - d\sigma^+}{d\sigma^- + d\sigma^-}
\\ \nonumber
A^{\rm pol}_{FB} &=&
  \frac{(d\sigma^-_{F} - d\sigma^-_B)-(d\sigma^+_F -d\sigma^+_B)}
{(d \sigma^-_{F} + d \sigma^-_B) + (d \sigma^+_F +d \sigma^+_B)}
\end{eqnarray}
Here 
$A^{\rm pol}_{FB}$
is the polarized forward backward asymmetry.
Fig.~\ref{fig:lhcbsm} shows the predictions of different models 
for $Z'$ in the $A_{LR}- A^{\rm pol}_{FB}$ plane. 
The LHC error bars correspond 
to 1000 $e^+e^-$ and 1000 $\mu^+\mu^-$ events, with 100\% polarization.
Obviously, it will be very easy to separate the various models with these
variables~\cite{lhcvirey}.

For new physics channels in the hadronic final states, one expects the 
reach to be larger for polarized compared to non-polarized measurements,
as for RHIC. Studies for the LHC suggest a possible improvement
of a factor of two for e.g. contact interactions or leptophobic 
$Z'$s~\cite{lhcvirey}.

At a high energy  $ep$-collider the region of high $Q^2$ offers the 
largest chance of discovering new physics.
Such new physics could manifest itself through the production
of new particles, such as Leptoquarks or SUSY particles in RP
violating models, contact interactions, etc.
A general study was made based on the 
contact interaction formalism~\cite{virey}. 
It was 
demonstrated that a fully polarized HERA would be very instrumental 
in disentangling the chiral structure of the new interactions.
With 250 pb$^{-1}$ data samples for  polarized $e^+$ and
$e^-$ beams, for each of the 2 spin orientations,
the asymmetries are sensitive to
contact interactions to scales
larger than 7 TeV (95\% C.L.).
In the presence of a signal these different
combinations of cross sections into the seven different asymmetries
allow a complete identification of the chiral structure of the new
interactions, i.e.~whether the interactions are LL, RR, LR or LR or 
a combination of those (where L and R denote the left and right handed 
fermion helicities  for the lepton and quark respectively).

This study has been further extended to the special case of 
leptoquark-like production~\cite{virey2}: asymmetries like the ones above
would allow one to pin down the chiral properties of the couplings 
to these new
particles.

These high $Q^2$ studies may be most relevant for HERA and THERA.
For THERA both $e^+$ and $e^-$ beams will be needed and
minimum integrated luminosities of order 100-200 pb$^{-1}$/beam.

In conclusion the studies for RHIC, LHC and HERA show that 
for the leptonic channels, polarization in general does not help 
to improve the search limits. In hadronic channels however, the 
analysis of spin asymmetries may help to considerably extend the 
search reach. In all cases polarization will be instrumental to 
analyse the chiral structure of new dynamics, if discovered.

\section{FUTURE FACILITIES}

The study of the spin structure of the proton and high-energy spin
phenomena has only just begun!

There are many urgent and open questions in spin physics requiring 
dedicated experimental and theoretical input.
Several newly proposed avenues of exploration, such as generalized 
parton distributions and  transversity studies, will be only barely 
touched upon by the presently planned experiments in the next few 
years.
The first measurements on $\Delta g$, flavour-separation, DVCS,
transversity... 
will be made by HERMES, RHIC-SPIN, COMPASS and 
experiments at Jefferson Laboratory and SLAC in the next five years
--- see Table 4.
However, it is clear even now that to really pin down the underlying 
physics, measurements in a larger kinematical range and/or with more 
precision will be required.
Only with a strongly focused and united position can 
the spin community be successful in keeping such a programme developing.

From the high energy end, what are the options ?
The HERA accelerator may match the time schedule best. 
The HERA luminosity has been upgraded and is expected to be around 
200 pb$^{-1}$/year/experiment.
HERA is now starting its phase II.
It has recently been decided that this programme will run up to and
including 2006. 
After that there may be room for a HERA phase III, depending on the 
strength of the physics programme, the community interested to carry 
such a program through, and the schedule of TESLA.

Polarized protons in HERA is a possibility for HERA III.
HERA already has a polarized electron beam, 
which will be used from
now by the collider experiments H1 and ZEUS, as well. 
These experiments could probably be continued 
to be used for polarized measurements, 
unless detector aging issues would be preventive.

The experience of RHIC, which starts to be commissioned as a polarized 
$pp$ collider in 2001, will be of vital importance. 
If successful, high energy polarized beams could become a `standard' 
for proton machines. 
To be useful for experiments one should be able to have stable proton 
beams with 
a polarization of larger than 50\%. 
Present estimates for HERA and RHIC 
are 
for a polarization of 70\%. 
An important issue is the measurement of the polarization of such a 
proton beam. At BNL the CNI mechanism in $pC$ scattering is presently 
exploited for this purpose~\cite{kurita}.

Preliminary price estimates for HERA are reported in~\cite{snowmass} and 
amount to 30M Euro to polarize HERA, possibly cheaper for deuterons.
Hence, it looks like this option can be realized
for a reasonable price bargain.
One could then pursue polarized deep inelastic scattering in a totally 
new regime sometime in the second half of this decade.
Since it could be technically simpler to have polarized 
deuteron beams instead of proton beams,
it is opportune that some of the key physics studies will be repeated 
for deuterons.
The machine group should however judge whether this initial enthusiasm 
can be backed up by calculations, like was done for polarized protons.

Continuing the HERA program would also encourage HERMES to plan 
further beyond 
their present aims. 
Ideas exist~\cite{nowak} to study the possibility of higher density 
polarized targets to better tackle e.g. DVCS and transversity measurements.

At RHIC possible upgrades include increasing the luminosity for $pp$
collisions by a factor 25-40 and upgrading the beam energy from 
250 GeV to 325 GeV. 
With such a collider, more subtle spin effects such as transversity 
in double spin asymmetries could become practicable.

The idea to add an electron accelerator onto RHIC is catching 
on~\cite{deshpande}. 
Two possible solutions are being investigated: 
adding an electron ring 
and 
adding an electron LINAC with an energy of 10 GeV
and with an intensity such that high luminosities, 
more than a factor of 100 larger than at HERA, can be achieved.
The polarization of the electron beam will be in the range of 
70-80\%. 
Note that the energy of the proton beam can be varied from 25 GeV to 
250 GeV, hence collisions with a CMS energy in the range of 10-100 GeV
are possible at RHIC.
If this scenario will be realized then a new experiment would 
be needed, carefully adapted to the 
interaction region and well integrated with the beam line magnets, 
allowing for $ep$, $eA$ and $pA$ collisions. 
This scenario offers the opportunity
to built a real optimized detector for such type of physics, with
e.g. good particle identification, and in 
particular paying special attention to the beam-pipe regions~\cite{krasny}.

The program of this Electron Ion Collider (EIC) or e-RHIC is very rich, and 
much of the physics potential can be appreciated  from the polarized HERA 
physics studies. 
The high luminosity is a clear advantage. 
In case polarized HERA will not happen, 
one should also 
seriously consider what the highest possible energy is that could 
be reached with the EIC. 
For example, with a 30 GeV electron beam the reach in $x$ and 
$Q^2$ is less than a factor 3 smaller than for 
HERA.
The status of the EIC project is that an R\&D program is being 
pursued in order  to have a proposal ready by about 2004. 
If all goes well $ep$ physics could be studied at BNL by about 2010.
If funding can be found for such a program, EIC may turn out to be 
the most attractive option in terms of expected performance and time-scale.

Possible projects beyond this time-line include those associated with an 
$e^+e^-$ linear collider which may come into operation 
some time after 2010. 
If such a collider were to be built close to a Laboratory with a 
high energy proton beam {\it and} if this proton beam is polarized, 
then one can push the frontier of DIS. 
The THERA studies have briefly looked into this option. 
Otherwise one can use the high intensity electron beam in fixed target 
mode, like TESLA-N or ELFE@DESY.
Uncanny precision measurements could be obtained at such a facility.
However one should note that if there were to be no continuation of a 
spin program beyond already planned and approved experiments, 
then the experimental side of this field may stop around 2006, 
leaving too long a gap before the start-up of these far-future facilities...

At intermediate energy, the proposed upgrade of Jefferson Laboratory 
to 12 GeV would provide continuous electron beams at high luminosity 
enabling precision studies of the valence structure of the nucleon and 
QCD ``confinement dynamics''.
A definitive $\nu p$ elastic experiment 
(as suggested for miniBooNE at FNAL~\cite{tayloe}) would provide a 
complementary window on the spin structure of the proton and enable
one to make an independent measurement of $g_A^{(0)}$ (including any 
possible contribution from $x=0$).


Building on the programme of polarized proton-proton collisions presently 
underway at RHIC it is worthwhile to investigate the physics potential of 
future polarized proton-proton collisions in the Large Hadron Collider, 
LHC, at CERN.
High-energy polarized proton-proton collisions could be particularly 
useful to probe the chiral and spin structure of supersymmetric 
couplings and extensions to the minimal Standard Model with an elementary 
Higgs sector which might be discovered with the (unpolarized) LHC,
including any possible strong (dynamical) electroweak symmetry breaking.
Certainly the LHC will be a collider facility for a long time to come
with e.g. luminosity and perhaps energy upgrades still to be planned.
The new physics may dictate that the use of polarized proton beams would 
be a vital discriminator between models and theories. Hence the option to 
polarize the LHC in future should be kept open.

In summary, the workshop led us to believe that the spin community 
has sufficient dynamics (and youth) to believe in the future of a 
long range program, in a similar way that NASA has set itself 
different landmark experiments for the next 20 years~\cite{snowmass2}. 
The community should not lose this energy, converge on a solid 
physics program and bring up the strength to see to it that it happens.

\begin{table*}[htb]
\caption{Key observables and experiments which will measure them}
\label{table:1}
\newcommand{\m}{\hphantom{$-$}}
\newcommand{\cc}[1]{\multicolumn{1}{c}{#1}}
\renewcommand{\tabcolsep}{2pc} 
\renewcommand{\arraystretch}{1.2} 
\begin{tabular}{@{}lll}
\hline
Quantity                  & Experiment & Dates      \\
\hline
$\Delta g$                & COMPASS    & 2003       \\
                          & HERMES     & 2002 +   \\
                          & RHIC       & 2002-05       \\
		          & SLAC E-161 & 2005  \\
Flavour separation        & HERMES     & 2002       \\
			  & RHIC       & 2002-04      \\
$\Lambda$ polarization    & RHIC       & 2002+      \\
Transversity, $h_1$       & COMPASS & 2004+ \\
                          & HERMES     & 2002-03       \\
			  & RHIC       & 2002+       \\
Transversity from $e^+ e^-$ & BELLE      & 2002       \\
DVCS plus meson production  & COMPASS    & 2004+      \\
			  & HERMES     & 2004-05      \\
High energy GDH integrand & SLAC E-159 & 2006         \\
\hline
\end{tabular}\\[2pt]
\end{table*}

\vspace{1.0cm}

We thank B. Badelek, P. Bosted, G. Contreras, A. Deshpande, F. Kunne, 
Z.-E. Meziani, W.-D. Nowak, G. R\"adel, J. Soffer and A.W. Thomas for 
helpful comments and communications.

\end{document}